**Marvin Steinke**

# Calibrating Microgrid Simulations for Energy-Aware Computing Systems

A thesis submitted to the
**Faculty of Electrical Engineering and Computer Science**
of the
**Technical University of Berlin**
in partial fulfillment of the requirements for the degree
**Master of Computer Science**

Advisor:                    Philipp Wiesner
First Reviewer:        Prof. Dr. habil. Odej Kao
Second Reviewer:   Prof. Dr. rer. nat. Volker Markl

Berlin, Germany
December 10, 2024

# Eidesstattliche Erklärung / Statutory Declaration

Hiermit versichere ich, dass ich die vorliegende Arbeit eigenständig ohne Hilfe Dritter und ausschließlich unter Verwendung der aufgeführten Quellen und Hilfsmittel angefertigt habe. Alle Stellen die den benutzten Quellen und Hilfsmitteln unverändert oder sinngemäß entnommen sind, habe ich als solche kenntlich gemacht. Sofern generative KI-Tools verwendet wurden, habe ich Produktnamen, Hersteller, die jeweils verwendete Softwareversion und die jeweiligen Einsatzzwecke (z.B. sprachliche Überprüfung und Verbesserung der Texte, systematische Recherche) benannt. Ich verantworte die Auswahl, die Übernahme und sämtliche Ergebnisse des von mir verwendeten KI-generierten Outputs vollumfänglich selbst. Die Satzung zur Sicherung guter wissenschaftlicher Praxis an der TU Berlin vom 8. März 2017[1] habe ich zur Kenntnis genommen.

I hereby declare that the thesis submitted is my own, unaided work, completed without any external help. Only the sources and resources listed were used. All passages taken from the sources and aids used, either unchanged or paraphrased, have been marked as such. Where generative AI tools were used, I have indicated the product name, manufacturer, the software version used, as well as the respective purpose (e.g. checking and improving language in the texts, systematic

---

[1] https://www.static.tu.berlin/fileadmin/www/10000060/FSC/Promotion___ Habilitation/Dokumente/Grundsaetze_gute_wissenschaftliche_Praxis_2017.pdf

research). I am fully responsible for the selection, adoption, and all results of the AI-generated output I use. I have taken note of the Principles for Ensuring Good Research Practice at TU Berlin dated 8 March 2017[2]. I further declare that I have not submitted the thesis in the same or similar form to any other examination authority.

Berlin, December 10, 2024

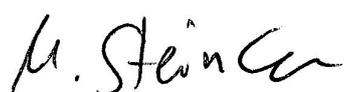
......................................................
Marvin Steinke

---

## Zusammenfassung

Die steigende Nachfrage nach Rechenressourcen erhöht den weltweiten Stromverbrauch in Rechenzentren, der bis 2026 voraussichtlich 1000 TWh überschreiten wird, was hauptsächlich auf die Einführung neuer KI-Technologien zurückzuführen ist. Kohlenstoffbewusste Strategien können Auswirkungen auf die Umwelt verringern, indem sie den Stromverbrauch mit der Produktion erneuerbarer Energien abstimmen, stehen jedoch vor Herausforderungen aufgrund mangelnder Entwicklungsumgebungen. Bestehende Lösungen basieren entweder auf teuren und komplexen physischen Systemarchitekturen, die schwer zu integrieren und zu warten sind, oder auf vollständigen Simulationen, die zwar kostengünstiger, aber oft unrealistisch sind, da sie System-Overheads und den Echtzeit-Energieverbrauch von Rechenknoten sowie Ressourcenschwankungen ignorieren.

Hierfür präsentiert diese Masterarbeit ein selbst-kalibrierendes, energiebewusstes Software-Testbed, das das Software-in-the-Loop Co-Simulations-Framework Vessim verwendet, um gleichzeitig Simulatoren für erneuerbare Energien und reale Knoten zu benutzen. Der Energieverbrauch auf Anwendungsebene dieser Knoten wird zunächst durch das Kepler Framework approximiert und dann in Vessims Microgrid-Simulation mithilfe eines externen Messgeräts auf Systemebene kalibriert. Die Simulation der teuersten und komplexesten Komponenten mit Einbindung realer Knoten stellt eine kosteneffiziente und realistische Lösung für energie- und kohlenstoffbewusste Systeme dar.

Die Auswertung des Testbeds mit intensiven GPU- und CPU-Arbeitslasten zeigt eine ziemlich genaue Leistungsapproximation des gesamten Knotens durch Kepler mit einem durchschnittlichen Regressionskoeffizienten von 1,01 und einem $R^2$-Wert von 0,95, obwohl bestimmte maschinelle Lernaufgaben eine höhere Abweichung zeigten. Der durchschnittliche statische y-Achsenabschnitt der Regressionslinie von ~5,23 W weist auf Ungenauigkeiten bei der approximierten Leerlaufleistung hin. Die Kalibrierung des dynamischen prozessbezogenen Energieverbrauchs verbesserte die Genauigkeit für GPU-Arbeitslasten um ~50 %, während CPU-Arbeitslasten eine geringe Verbesserung von ~3,5 % verzeichneten.

# Abstract

The surge for computing resource demand is increasing global electricity consumption in data centers which is expected to exceed 1000 TWh by 2026, mainly attributable to adoption of new AI technologies. Carbon-aware computing strategies can mitigate their environmental impact by aligning power consumption with the production of low-carbon renewable energy, but they face challenges due to the scarcity of development environments. Existing solutions either rely on costly and complex physical system architectures that are difficult to integrate and maintain or on full simulations that, while more economical, often lack realism by ignoring system overheads, and real-time node power consumption and resource fluctuations.

This thesis remediates these issues by proposing a self-calibrating energy-aware software testbed that uses the Software-in-the-Loop co-simulation framework Vessim to integrate renewable energy production simulators, while including real computing nodes. The application-level power consumption of these are first approximated by the Kepler framework and then calibrated within Vessim's microgrid simulation using an external socket power meter as a definitive measurement source on the system-level. Through the simulation of the most costly and complex components and the inclusion of real computing nodes, the testbed provides a cost-effective yet realistic solution for energy- and carbon-aware computing systems.

The evaluation of the testbed with GPU and CPU intensive workloads reveal fairly accurate power approximation of the whole computing node by the Kepler framework, with an average regression coefficient of 1.01 and $R^2$ values of 0.95, though certain machine learning workloads showed higher deviation. The average static y-intercept of the regression line of ~5.23 W indicate inaccuracies in the idle power approximation. Calibration of dynamic per-process power consumption improved accuracy for GPU workloads by ~50%, while CPU workloads saw a modest improvement of ~3.5%.

# Contents

# Chapter 1

# Introduction

The exponentially increasing need for computing resources [1] and, in turn, the expansion of cloud platforms is significantly driving up global electricity demand. In 2022, it was estimated that data centers around the world consumed between 240 and 340 terawatt-hours (TWh) of electricity, an increase of 20% to 70% compared to 2015. This accounts for approximately 1 to 1.3% of the global final electricity demand [2]. Recent advancements in machine learning (ML), particularly with technologies like the Generative Pre-trained Transformer (GPT) models [3], also currently play an important role in this trend. These models are generally trained using considerable computational resources [4]. While the carbon emissions associated with the inference operations of such models remain largely unquantified [5], they are potentially substantial [6]; projections suggest that total electricity consumption by data centers could surpass 1000 TWh by 2026 [7], more than doubling their current usage, primarily due to the increased deployment of AI technologies.

As environmental concerns parallel the growing energy demands associated with computing infrastructures, data centers have become notable contributors to global greenhouse gas (GHG) emissions, accounting for around 330 million tonnes of $CO_2$ equivalent in 2020 alone [8]. This represents approximately 1% of energy-related emissions worldwide [9]. Despite achieving near-optimal levels in energy efficiency recently [10], further reductions in emissions are impera-tive. The adoption of carbon-aware computing strategies, which align the energy consumption of computing systems with the availability of intermittent low-carbon energy sources such as solar and wind, can significantly mitigate environmental impact [11]–[15].

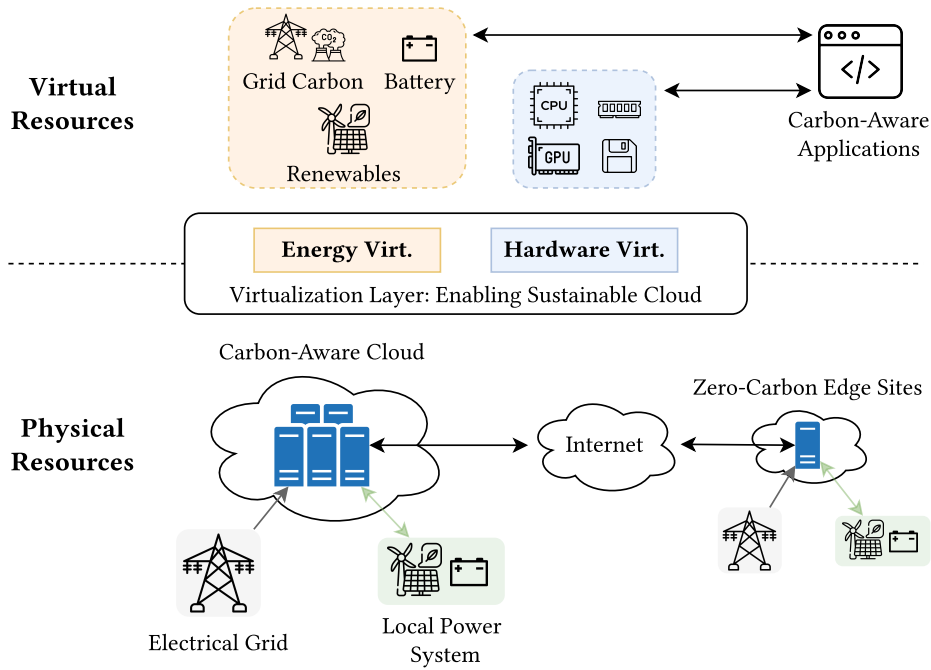

Figure 1: Sustainable Cloud-Edge Infrastructure Vision[3]

These strategies are feasible since computation often exhibits great flexibility in spatial, temporal, and performance parameters, facilitating execution relocation, timing, and intensity [17]. Bashir et al. [16] advocate for such alignment at an application-level, arguing for the virtualization of the power system to provide visibility and software-defined control over energy resources. Exposing the power system to cloud applications would enable them to define their own energy and carbon emission management strategies, optimizing energy consumption and reducing carbon emissions within a virtualized cloud-edge infrastructure, as illustrated in Figure 1. Souza et al. [13] implement this vision with *ecovisors*, which monitor power by tracking generation and consumption via APIs and measure real-time carbon inten-sity using services like Electricity Maps[4] and WattTime[5]. They regulate server consumption and battery cycles based on carbon intensity and energy availability, capping power usage per container and managing batteries via APIs, as shown in Figure 2.

---

[3]Adapted from Bashir et al. [16]: Abstracted infrastructure and improved clarity.

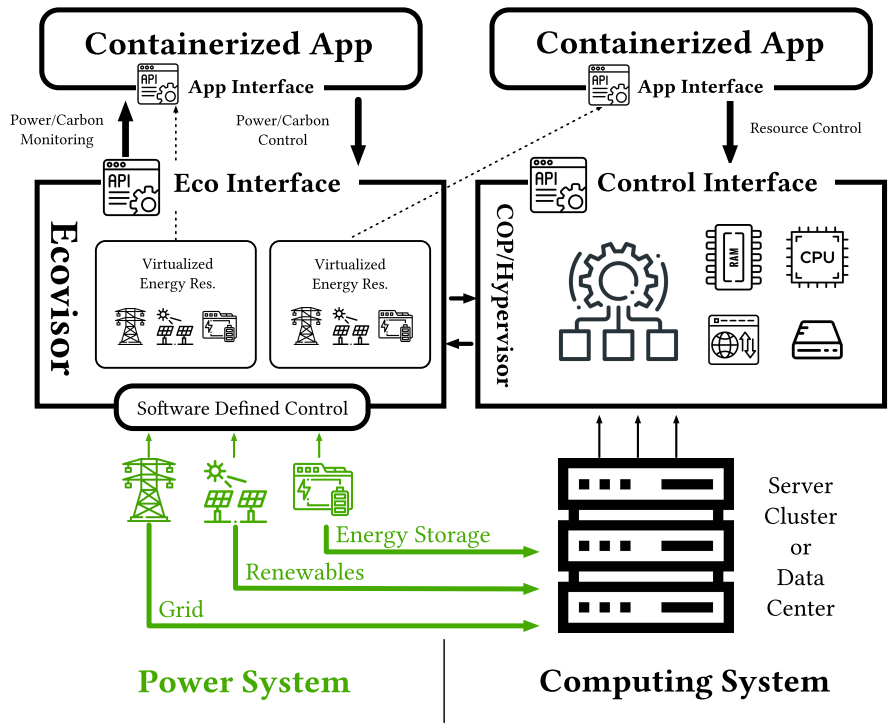

Figure 2: Ecovisor Design[6]

Research and development of carbon-aware applications and systems, however, continue to face challenges due to the limited availability of testing environments. Souza et al. [13] implemented an ecovisor on a hardware testbed, including a few low-powered microservers, but even this small-scale physical prototype proved to be both very complex and costly, primarily due to the expensive components of the power system and their intricate integration into the system. Consequently the scalability of this testbed is limited. Moreover, this prototype only measures the power consumption of each server node and is not capable of power measurements for an individual application. While hardware testbeds are important for real-world testing, their high costs and maintenance requirements make them feasible only for those with substantial resources [11].

---

[4]Electricity Maps. https://www.electricitymaps.com/

[5]WattTime. https://watttime.org/

[6]Adapted from Souza et al. [13]: Replaced LXD with generic COP/hypervisor depiction, renamed and colored the ("energy" → "power") system.

As an alternative, full simulations offer a viable and more economical approach for testing and developing these systems. However, these simulations frequently suffer from a lack of realism. Their approximations of nodes' runtime behaviors typically fail to factor in real-time fluctuations in energy and resource utilization and operate on a discrete time basis, ignoring the impacts of system overheads [15], [18]–[21]. Wiesner et al. address these issues with *Vessim*, a Software in the Loop (SiL) co-simulation testbed for carbon-aware applications and systems [22], [23]. Given that power systems in these testbeds are often the most complex and costly components to integrate, they simulate only the microgrid portion of such a system, denoted in green in Figure 2, while the applications continue to run on real hardware. This approach allows overheads and fluctuations to be incorporated into the simulation, increasing its degree of realism. Yet, their particular design requires applications to measure their own power consumption and communicate this data to the simulation environment through a message broker, resulting in a testbed that is virtually detached from the computing system.

Including online, attributable power consumption of these applications is currently only possible through approximation with a combination of process-related resource utilization metrics and multiple hardware counters and sensors [24]. Numerous studies validate the accuracy of power meters for main components of computing systems, the CPU, memory, and GPU. For CPU and DRAM power consumption, Intel's Running Average Power Limit (RAPL) [25] features a ~1ms [26] sampling interval, while demonstrating very low error margins across various models, with a workload-dependent accuracy of ± 2W at the 95th percentile [27], [28]. In contrast, NVIDIA GPU power measurements, which utilize more primitive onboard sensors, exhibit significant variability between models. Recent GPUs typically have a sam-pling rate of around 20ms [29] and maintain a percentage error within ± 5% [30], which challenges NVIDIA's claim of a ± 5W deviation [31].

These error margins for each component accumulate over time, exacerbating inaccuracies in estimating power consumption. Margins vary depending on the type of workload, which also causes further inaccuracies when attributing power consumption to individual applications based on resource utilization. Different types of workloads can result in varying power consumption despite similar resource utilization due to the distinct operations they perform on the hardware [32]. Consequently, the result is not an accurate representation of an application's power consumption. Research in this domain has predominantly focused on calibrating power models for public cloud environments, where direct hardware power measurements are (practically) not accessible [24], [33]–[35]. Despite the progress, there has been no significant effort to calibrate these hardware measurements against a verified standard at runtime, which would benefit applications that rely on accurate microgrid simulation behaviors.

## 1.1 Testbed Requirements

As stated above, current testbeds frequently suffer from ample complexity and cost, insufficient scalability, lack of realism, or inadequate implementation of ecovisor principles, critical for the research and development of carbon-aware applications and systems. To validate application power consumption approximations and accurately calibrate microgrid simulations with a source of truth for total system power consumption, a testbed featuring the following properties is necessary:

**(1) Ecovisor Principles**  Souza et al [13] define the properties of an ecovisor through three main capabilities. (1.1) It is proficient at monitoring power generation and consumption from various energy sources such as the grid, batteries, and renewables, as well as server power consumption. (1.2) It can monitor the carbon intensity of grid power in real time. (1.3) The ecovisor can regulate power usage, including server power consumption and battery charge/discharge, based on carbon intensity and the availability of renewable energy.

**(2) Orchestration of Applications** Applications must be managed non-intrusively, i.e. they should not necessitate modifications to their implementation (2.1) or be required to report their own power consumption (2.2). Either a hypervisor (VMM) for virtual machines (VMs) or a container orchestration platform (COP) for containers is suitable to coordinate the applications for communication, control, and management across the network.

**(3) In-Place Vertical Scalability** Since power consumption on an application-level is controlled through resource caps, applications need to be variable in their individual computing capabilities. In-place scalability denotes that an application's resources can be dynamically adjusted at runtime, i.e. without requiring a restart.

**(4) Economical Viability** The testbed should be (4.1) inexpensive and (4.2) simple to set up and maintain. Because the power system is the most complex and costly component to integrate, it should be simulated rather than implemented physically.

## 1.2 Contributions

The proposed energy-aware software testbed represents an advancement in reducing development costs, time, and complexity for scalable carbon-aware cloud computing systems. By integrating application-level power approximations and system-level power measurements, the testbed improves the realism of microgrid software-in-the-loop (SiL) simulations. This is achieved through an online calibration approach that fine-tunes application-specific power consumption models for incorporation into Vessim.

The testbed assesses the effectiveness of contemporary hardware sensors and power models to accurately measure and approximate computing system power consumption. The open-source Kubernetes-native framework, Kepler, is validated for its ability to integrate these sensors and models to estimate each process's power consumption.

The findings and documentation from this process are contributed to aid in the framework's development. Additionally, the testbed supplements Vessim's system design with the option of a COP for application orchestration, enabling the interaction of SIL applications and systems with the simulation, as originally outlined by Wiesner et al. [23]. Throughout the benchmarking process in evaluating the testbed's ability to calibrate power consumption, multiple applications are subject to the microgrid calibration procedure as an abstract representation within Vessim. The specific implementations of these applications are demonstrated, providing valuable insights that aid in the configuration of similar evaluation setups for researchers.

## 1.3 Outline

The remainder of this thesis is structured as follows:

Chapter 2 provides background information on discussed subjects to establish a foundation for the thesis. First, as the Vessim simulator finds extensive application in this thesis, the topics simulation modeling, validation, and the primary method utilized in the approach, calibration, are introduces. Second, microgrids, as localized power systems, are explored and Vessim is introduced, a microgrid simulation testbed that forms the basis of the energy-aware testbed. Third, power consumption measurement techniques in computer hardware are presented, with a particular emphasis on the Kepler framework that is used to estimate per-process power consumption and represents a key component of the testbed architecture.

Chapter 3 reviews literature that is related to this thesis categorized, as 1. energy-aware computing system, 2. attribution of power and carbon consumption, and 3. calibration of power systems. Each section represents similar research efforts to different concepts revolving around the testbed and its applications, while no approach solves all the challenges that are outlined in Section 1.1.

Chapter 4 introduces the energy-aware testbed. First, its architecture is described, with its individual main components and their interconnections: applications, monitoring, simulation environment, and resource control forming a feedback loop. Second, the specific Kubernetes implementation is detailed, packaged as a Helm chart and third, the hardware and software requirements are outlined for users to install and operate the testbed.

Chapter 5 demonstrates the calibration procedure that is employed by the testbed within Vessim's simulation scenario. First, application-level power consumption approximation and system-level socket-meter power measurements are used for online-calibration of idle and dynamic per-process power. Second, the implementation of this procedure is into Vessim's microgrid simulation is detailed.

Chapter 6 evaluates the testbed and its calibration methodology. First, the benchmarking setup is described, including details to the composition of the host system, the overall benchmark design, and its implementation and execution in Vessim. Second, the workloads used in the benchmarks are introduced, and their specific configurations are detailed. Third, the results of the benchmarks are presented and analyzed. This section is partitioned into the validation of total system power approximations and the calibration of per-process power consumption. Additionally, any restriction that may interfere with the testbed's operation are discussed. Fourth, and finally, the conformity of the testbed with the requirements as defined in Section 1.1 is examined.

Chapter 7 concludes this thesis by summarizing its main points and contributions and outlining future work related to the testbed and its per-process calibration ability.

# Chapter 2

# Background

This chapter covers essential concepts for understanding simulation modeling, microgrids, power systems and measurement of power consumption in computing systems. Section 2.1 introduces simulation modeling as a method for representing complex systems to predict outcomes, including the need for validation and calibration to ensure model accuracy. Section 2.2 explores microgrids, localized energy systems which integrate with the main power grid or operating independently. This section also introduces Vessim, a testbed for simulating microgrids. Section 2.3 discusses power consumption measurement, highlighting tools like Intel's RAPL and frameworks like Kepler to optimize energy use and reduce carbon emissions in computing systems.

## 2.1 Simulation Modelling

Simulation modeling is a computational technique for creating simplified representations of real-world systems to explore their behavior, predict outcomes, and support decision-making. It involves identifying key system components and interactions, setting assumptions often in the form of mathematical or logical relationships, and constructing a model. While mathematical methods can solve some simple models analytically, most systems are too complex for this, requiring numerical evaluation through simulation [36]. To ensure that a simulation model truly reflects the complex realities it aims to replicate, validation and calibration are necessary.

### 2.1.1 Validation

Validation is the process of verifying that a simulation model accurately represents the real-world system. According to Law et al., a model is considered *valid* if it can facilitate decision-making about the system in a manner comparable to conducting experiments on the actual system if it were feasible [37]. To determine the validity of a simulation model, the most definitive test is to ensure that its output data closely match the expected output data from the actual or proposed system. This process is known as results validation.

Regression analysis plays a crucial role in this context and this thesis, as it serves as a tool to model the relationship between dependent and independent variable to enable predictions and estimations of outcomes. Types of regression models include linear, multiple, and logistic regression, each chosen based on the specific characteristics and requirements of the data at hand. To assess how well the regression line fits the given data, metrics like $R^2$ and Adjusted $R^2$ are utilized, providing a measure of the model's explanatory power. Furthermore, residual analysis is an important step, examining the differences between observed and predicted values [38]. This analysis helps identify any patterns or anomalies, ensuring the validity of the regression model and guiding necessary adjustments for improved accuracy.

### 2.1.2 Calibration

Simulation models are typically designed for broad applicability. However, to generate accurate predictions for specific contexts, they often need calibration with observed data [39]. The process of model calibration involves systematically refining model parameters to reduce the differences between the model's predictions and observed real-world data. Occasionally, parameters are adjusted without strong theoretical rationale, a practice known as *tweaking*. This process continues until the model's output closely matches the observed data. This practice raises concerns about the model's validity, specifically whether the adjusted model is representative of the underlying system or merely overfitted to the calibration data. To address this issue, the model is to be tested with a separate, independent set of input and output data. By comparing the model's predictions on this new data to the system's actual output, one can assess whether the calibration is robust and generalizable beyond the original data set [37]. In addition for the context of this thesis, online simulation calibration enhances model accuracy by adjusting simulation parameters in real-time, using current data. Mathematical models and approximation techniques reconcile discrepancies between observed and simulated data to ensure the simulation remains applicable and valid amid changing conditions, proving especially useful in systems with dynamic variables.

## 2.2  Microgrids and Power Systems

Olivares et al. define a microgrid as a localized energy system that can operate independently or in conjunction with the main power grid [40]. They integrate Distributed Energy Resources (DERs), Energy Storage Systems (ESSs), and controllable loads, and are perceived by the main grid as a single entity responding to control signals. Composed of a cluster of loads, Distributed Generation (DG) units, and ESSs, microgrids are connected at a single Point of Common Coupling (PCC). This setup facilitates decentralized problem-solving and reduces the need for complex central coordination, thereby supporting the realization of the Smart Grid [41]. In this thesis, the term *power system* is used in place of *energy system*. The latter is commonly used in the literature to refer to multi-energy building systems that include both electrical and thermal energy, while the former is specific to electrical energy [42].

Data centers often contend with substantial challenges, including escalating electricity costs, increasing carbon footprints, and unexpected power outages, which can be effectively mitigated through the implementation of such microgrids. By incorporating within their own microgrids with renewable energy sources and opportune storage, data centers can enhance energy efficiency, boost sustainability, and improve the reliability of their electrical services [43].

### 2.2.1 Vessim

To enable cost-efficient development and testing of applications and systems associated with microgrids, Wiesner et al.'s Vessim serves as a co-simulation testbed that integrates discrete-event simulators for various aspects of power systems, facilitating the inclusion of real applications and hardware (SiL) for a realistic representation of microgrids for distributed computing systems [22]. In Figure 3, Vessim enables the parallel simulation of multiple microgrids, each comprising a set of defined components. Hexagons denote co-simulation subsystems managed by the integrated Mosaik framework that forms the basis for the orchestration of all simulators [44]. The system includes actors, a grid simulator, controllers, and ESS simulators, which are stepped in this order.

An actor $a \in A$ within a microgrid can be a power producer ($p_t^a > 0$) or consumer ($p_t^a < 0$), communicating their power production or consumption at simulation step $t \in N$ to the grid simulator and optionally sending additional state information $state_t^a$ to controllers. The default grid simulator in Vessim aggregates net power from all actors to determine the grid's current surplus or deficit $\Delta p_t$. Controllers allow users to monitor and manage microgrid states based on aggregated information on each simulation step, including $state_t^a \forall a \in A$, $\Delta p_t$, energy storage states $state_t^s$, energy surplus/deficit since the last step $e_{t-1}$, and other custom input signals. A controller can, for instance, save and visualize data, vertically resize compute nodes' resources, manage the ESS or provide API access to real applications through the built-in `SilController`. The ESS simulator $s$ handles the charging and discharging of energy storage by modeling time progression between simulation steps. They receive $\Delta p_t$ and the elapsed duration between $t-1$ and $t$ to determine the energy settlement with the ESS or the public grid based on user-defined policies.

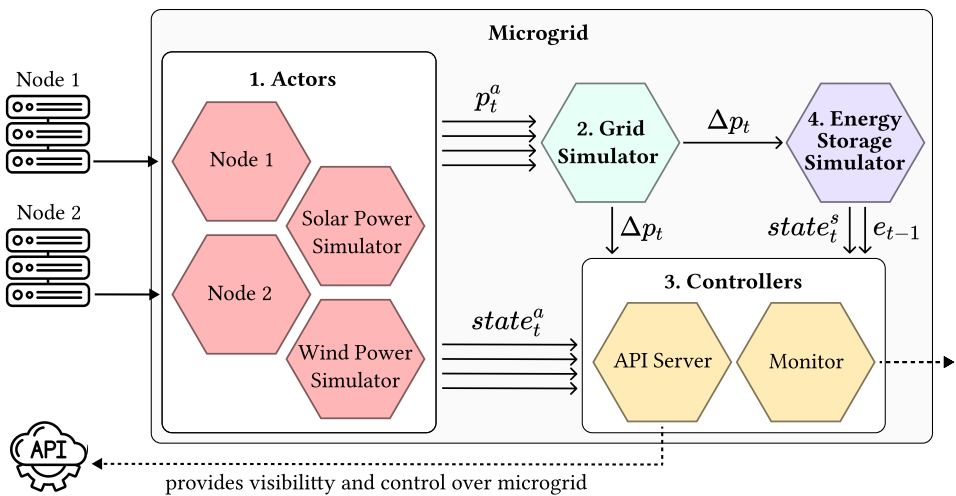

Figure 3: Vessim Simulation Scenario Architecture [22]

## 2.3 Measurement of Power Consumption

Power consumption is a critical metric in computing systems. Accurate measurements facilitate optimizing energy efficiency in software applications and reducing their carbon emissions by supporting the development of carbon-aware computing strategies, and allowing researchers to identify areas of excessive energy usage to make targeted improvements. Power meters are used to measure consumption at various levels of the system, ranging from basic platform sensors, monitoring components like the power supply unit (PSU), mainboard or fans, to more advanced meters dedicated to major power-consuming components such as the CPU, GPU, or memory.

### 2.3.1 Main Component Power Meters

Running Average Power Limit (RAPL) is an interface developed by Intel for enforcing power limits on CPUs and memory subsystems in real-time. Using a combination of hardware counters and calibrated power models, RAPL measures energy utilization and dynamically adjusts power states to maintain the desired power envelope [25].

RAPL supports multiple power domains, each corresponding to physical components for power management. Khan et al. depict the hierarchy of these power domains in Figure 4 [45]. The *Package* (PKG) domain encompasses the energy usage of the entire socket, including all cores, integrated graphics, and uncore components such as last-level caches and memory controllers. The *Power Plane 0* (PP0) domain specifically measures the energy consumption of all processor cores within the socket. In contrast, the *Power Plane 1* (PP1) domain gauges the energy usage of the processor graphics (GPU), albeit being specific to desktop models. The *DRAM* domain is responsible for measuring the energy consumption of random access memory (RAM) attached to the integrated memory controller. The *PSys* domain monitors and controls the thermal and power characteristics of the entire System on Chip (SoC), making it particularly valuable for tracking power consumption sources beyond the CPU and GPU, including the package domain, System Agent, PCH, eDRAM, and additional domains within a single-socket SoC.

NVIDIA GPUs have transitioned from using estimation-based to measurement-based methodologies for tracking power consumption. Older models, such as from the *Fermi* architecture from 2010 estimate power usage by monitoring activity signals. Newer GPUs from the *Kepler* architecture from 2012 onward, in contrast, utilize shunt resistors

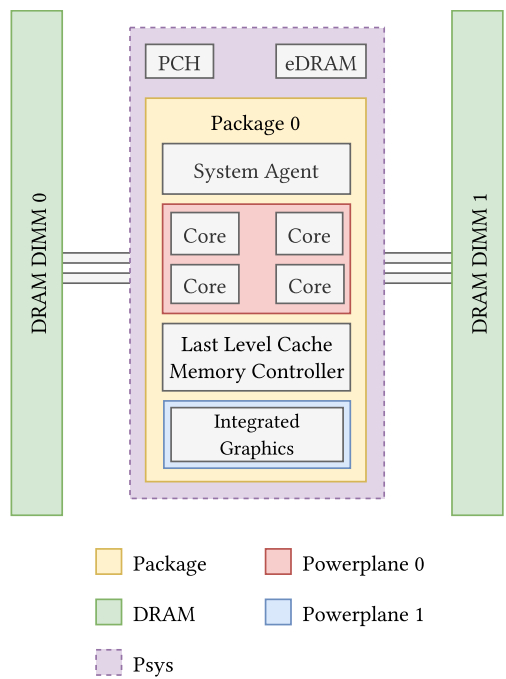

Figure 4: Power domains supported by RAPL[7]

---

and current monitor integrated circuits (ICs) for direct power measurements [30]. The NVIDIA System Management Interface (nvidia-smi), part of the NVIDIA Management Library (NVML), provides a command-line utility for monitoring, managing, and querying various GPU statistics [31].

### 2.3.2 Platform Power Meters

The Advanced Configuration and Power Interface (APCI) can be used to measure power consumption by accessing power sensor data available in the system's hwmon in sysfs [46]. The hwmon subsystem in the Linux kernel provides a standardized interface for hardware monitoring[8]. This interface allows users to query sensor data independently of specific hardware implementations by reading files located in the `/sys/class/hwmon` directory (sysfs). The values reported by these methods depend on the various hardware drivers in use. The Linux kernel scans available sensors and creates corresponding hwmon entries based on the detected hardware.

However, if available, the more recent Redfish (or the older Intelligent Power Management Interface (IPMI) [47]) standard can access more detailed power consumption data. Many modern hardware components, including storage devices, network interfaces, PSUs, or fans or other cooling units often have dedicated sensors or power metrics reported directly through the Redfish API [48]. Redfish works by utilizing a RESTful interface to deliver power monitoring data in a standardized JSON format and allows administrators to access real-time power consumption metrics, historical power usage data, and power thresholds.

### 2.3.3 Kepler Framework

Kepler, the Kubernetes-based Efficient Power Level Exporter framework, estimates power consumption at the process, container, and K8s pod levels [24]. It employs an extended Berkley Package Filter (eBPF)[9] program integrated into the kernel to extract resource utilization metrics. Kepler collects real-time power consumption metrics using the previously detailed APIs for power metering: RAPL and NVML for main components, and Redfish/IPMI and ACPI for platform-level measurements. When real-time metrics are unavailable, e.g. in VMs, Kepler utilizes regression-based trained power models. The architecture of Kepler is depicted in Figure 5.

---

[8]Linux hwmon Subsystem. https://hwmon.wiki.kernel.org/
[9]Extended Berkley Package Filter. https://ebpf.io/

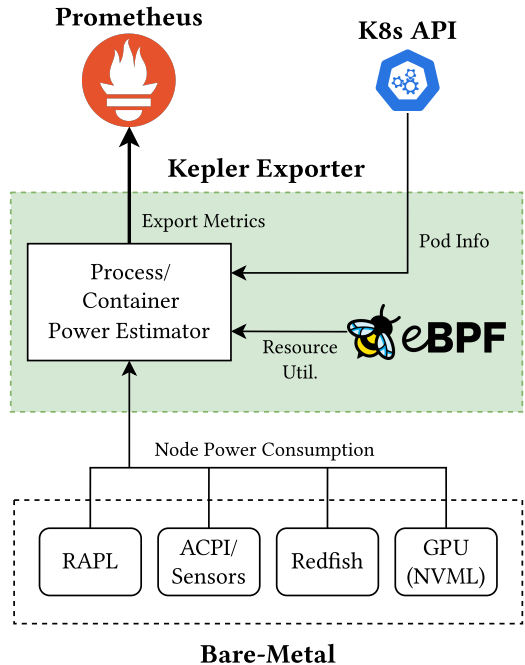

Figure 5: Kepler Bare-Metal Deployment Architecture

Kepler differentiates itself from existing power modeling frameworks such as Joulemeter [49], HaPPy [50], BitWatts [51], SmartWatts [33], and Power Containers [52] (with the latter three being based on Power-API [53]) by emphasizing per-process metrics rather than relying solely on aggregated system metrics. Unlike these approaches, which estimate an application's power consumption by summing its values across the system and comparing it with total system power, Kepler focuses on obtaining accurate power consumption data for individual applications. This is achieved by leveraging incremental power measurements in controlled environments to train and evaluate its power model. Furthermore, Kepler introduces a novel refinement mechanism that employs real-time system power metrics to improve accuracy. By normalizing estimated power with actual observed system power, Kepler aims to more precisely reflect real-world variables like ambient temperature changes, power state fluctuations, supplier inconsistencies, and hardware errors. This approach is particularly relevant given previous findings of substantial variations in power consumption of up to 30% in comparable nodes within a datacenter and up to 25% within the same node [54].

Kepler identifies the container a process belongs to using the Process ID (PID) and container ID through Linux cgroups[10]. It relies on K8s' API server to maintain an updated list of pods and classifies non-K8s container processes as *system processes*. To calculate energy consumption, Kepler employs its Ratio Power Model, which divides power usage based on the process-system resource utilization ratio. The APIs provided expose absolute power metrics that encompass both dynamic and idle power [55].

Dynamic power is associated with resource utilization, whereas idle power remains constant regardless of the system load. The dynamic and idle power components are allocated differently across processes. Let $P$ represent a set containing all processes of a worker node $n$. The Ratio Power Model divides dynamic power consumption $c^{\mathrm{dyn}}$ by calculating the ratio of a process's $p \in P$ resource utilization $r_p^{\mathrm{util}}$ to the total system utilization and then multiplying this ratio by the total dynamic power consumption to estimate the power usage for each process:

$$c_p^{\mathrm{dyn}} = \frac{c_n^{\mathrm{dyn}} \cdot r_p^{\mathrm{util}}}{\sum_{q \in P} r_q^{\mathrm{util}}}$$

This approach ensures the power usage estimates are closely based on actual utilization. Different types of resource utilizations are estimated distinctly: CPU utilization is assessed using CPU instructions, memory utilization through cache misses, and GPU utilization by

---

[10]Linux cgroups. https://docs.kernel.org/admin-guide/cgroup-v1/cgroups.html

evaluating Streaming Multiprocessor (SM) utilization. When the total resource usage is zero, power is evenly divided among processes as a fallback mechanism. Amaral et al. state, that the estimation of idle power adheres to guidelines from the GHG protocol [56], distributing idle power among processes according to their relative size, or requested resources $r^{\text{req}}$ [24]. However, this is currently determined by dividing the idle power by the number of processes instead, with the GHG protocol conformant estimation planned as a future work item[11]:

<div>

Stated Implemenation:

$$c_p^{\text{idle}} = \frac{c_n^{\text{idle}} \cdot r_p^{\text{req}}}{r_n^{\text{req}}}$$

Current Implemenation:

$$c_p^{\text{idle}} = \frac{c_n^{\text{idle}}}{|P|}$$

</div>

Kepler then exports these power metrics to Prometheus as a *counter*: a discrete-time, non-decreasing function $f : \mathbb{N} \to \mathbb{R}^+$ such that

$$\forall t_2, t_1 \in \mathbb{N}, t_2 \geq t_1 \Rightarrow f(t_2) \geq f(t_1)$$

The accumulated counter value within a specific interval is obtained by calculating the difference between the counter values at the end and the start of the interval. Due to potential fluctuations in counter metrics, a moving average might be opportune to determine the average power within the time interval:

$$\frac{f(t_2) - f(t_1)}{t_2 - t_1}$$

---

# Chapter 3

# Related Work

This chapter offers an overview of the state-of-the-art relevant to the thesis. Section 3.1 introduces systems that integrate decision-making capabilities centered around energy and carbon consumption. Section 3.2 outlines methods for attributing power and carbon consumption to tenants within multi-tenant cloud data centers or individual processes. Further, Section 3.3 explores systems designed to calibrate estimated power consumption in virtualized and distributed environments, often combining similar principles from sections 3.1 and 3.2.

## 3.1 Energy-Aware Computing Systems

Within 2023, energy-aware computing systems have gained popularity, particularly in conjunction with renewable energy and carbon intensity considerations. Bashir et al. and Souza et al. have been significant contributors to this trend. As outlined in Chapter 1, they argue for the virtualization of the energy system to expose software-defined control and visibility on an application-level [13], [16]. However, their prototype faces challenges in scalability, high costs, and a complex system architecture, making it difficult for researchers to adopt their approach as carbon-aware testbeds.

Hanafy et al. [57] adopt their principles of energy, carbon and resource monitoring but exercise a macro perspective with their *carbon scaling* approach. Aiming to mitigate carbon emissions within cloud computing, they focus on exploiting the elasticity of batch workloads by adjusting server allocations on a COP-level in response to variations in the carbon cost of energy. Their method revolves around a dynamic adjustment process, akin to cloud autoscaling, which optimizes carbon emissions by minimizing usage during high-carbon periods. Hanafy et al. implement a prototype, *CarbonScaler*, within K8s to utilize its autoscaling capabilities and an analytics tool to ensure carbon-efficient deployment of batch applications. While they monitor and control power consumption within their COP-based system and applications' architecture, their approach is highly intrusive and specific to batch workloads.

A less intrusive strategy is introduced by Qiao et al. [58] who enhance energy efficiency in the Linux operating system by optimizing the OS scheduler. To address the challenge of variable energy consumption across processes in datacenters, the authors propose modifications that do not require changes to the Linux kernel. They focus on two main challenges: gaining detailed energy insights per process with minimal overhead and developing energy-aware scheduling policies. *Wattmeter,* an energy measurement tool using eBPF and RAPL, measures energy use at the process level with low overhead by attaching eBPF programs to Linux's `sched_switch` tracepoint. For scheduling, they implement two policies via Google's ghOSt framework [59]: the Energy-Fair Scheduler (EFS), which balances energy consumption by adjusting process scheduling based on energy efficiency, and the Energy-Capped Scheduler (ECS), which monitors energy usage against set limits to prevent excess consumption. Qiao et al.'s monitoring approach is very similar to Amaral et al.'s [24] Kepler, as part of this thesis' prototype, but RAPL estimates are not read from the userspace as much finer grained sampling with less overhead is necessary for the process scheduler but no other hardware components are included. While no other hardware components are included, their constraint of this very fine-grained sampling also excludes the possibility of a more general approach to power monitoring.

Costero et al. [60] present an infrastructure aimed at maximizing performance in multi-application environments while adhering to power constraints. This system employs BAR (Power Budget-Aware Runtime Scheduler) and BARMAN (Power Budget-Aware Resource Manager). BAR is an extension of an OpenMP [61] runtime task scheduler that dynamically optimizes task-based application performance by employing techniques like dynamic voltage-frequency scaling (D-VFS) and thread malleability. It intelligently manages idle worker threads and core frequency to enhance performance within power limits without the need for user intervention. BARMAN operates at the system level to manage resources through dynamic power budget redistribution, ensuring that applications underutilizing their power allocation can relinquish resources to those that could better use them. The utilization of DVFS, however, is 1. limited to the CPU and disregards GPU involvement, and 2. can only target the entire CPU package, making the system design rigid for usecases that are not bound to a power budget.

## 3.2 Attribution of Power and Carbon Consumption

Accurate attribution of power and carbon consumption in multi-tenant data centers is important for sustainable operations and fair distribution of environmental impact amongst tenants. Various approaches are presented in this section to address challenges revolving around this topic, including allocation models for total carbon footprint, a Linux-based carbon-aware daemon, a game-theoretic method using the Shapley value, and a NUMA-aware model focusing on thread-level energy attribution accuracy. While none of these papers is directly related to the core approach of this thesis, many parallels can be drawn to power attribution framework Kepler that is used and validated extensively.

Westerhof et al. [62] discuss the development of an allocation model to attribute carbon emissions to tenants in multi-tenant cloud data centers. The model aims to calculate the Total Carbon Footprint (TCF) by considering operational emissions. The TCF calculation includes Scope 1, 2, and 3 emissions based on the GHG Protocol [56] but specifically addresses multi-tenancy challenges. Implementation details involve defining a responsibility ratio to fairly distribute emissions among tenants, utilizing existing methods to estimate server and network energy consumption, and using regression to model energy usage based on parameters like CPU and disk usage. The model also considers carbon intensity for each data center and accounts for offsetting measures such as purchasing Renewable Energy Certificates (RECs).

Schmidt et al. [63] present *carbond,* a Linux-based operating system daemon designed to enable carbon awareness by tracking carbon emissions at both operational and embodied levels. The carbond service measures energy consumption to calculate operational carbon emissions. It collects carbon intensity data from hardware components and provides an interface for application-level software to access this information, while following the Software Carbon Intensity (SCI)[12] standard and supporting fine-grained units of work. The authors acknowledge the possibility of integrating carbond with Souza et al.'s Ecovisor [13], which would alleviate problems of their current prototype with undisclosed power consumption sources and missing GPU involvement. An integration with existing COPs, adding carbon consumption as a monitoring resource comparable to CPU and memory, would be a valuable extension.

Han et al. [64] present a game-theoretic approach using the Shapley value to equitably attribute carbon emissions among workloads in cloud data centers, addressing deficiencies in traditional methods. The Shapley value ensures a fair distribution of emissions by considering principles such as null player, symmetry, efficiency, and linearity. It uses power and resource utilization data to calculate each workload's average marginal contribution, considering all possible workload arrangements. For operational carbon attribution, power consumption is assigned based on a workload's share of total system power over time, incorporating grid carbon intensity for footprint calculations. The model also covers embodied carbon by evaluating resource demand and peak provisioning, which includes components like CPUs and DRAM. It distinguishes between fixed and variable embodied carbon to allocate emissions based only on the actual resources utilized.

---

[12]Software Carbon Intensity (SCI) Specification. https://sci.greensoftware. foundation/

Hè et al. [65] address the accurate attribution of energy consumption in multi-tenant computing environments, emphasizing the inadequacy of traditional coarse-grained models, particularly in Non-Uniform Memory Access (NUMA) architectures. The authors propose an approach with a thread-level, NUMA-aware model that improves energy attribution accuracy, leveraging the Intel Running Average Power Limit (RAPL) meter for initial measurements. The model integrates an energy crediting strategy, mitigating the noisy-neighbor effect by correlating energy usage with actual resource consumption instead of total resource usage. The implementation, termed EnergAt, functions as a separate process that periodically samples CPU and memory statistics every 10 milliseconds, enabling precise calculation of energy consumption per thread and socket.

## 3.3 Calibration of Power Systems

This section presents approaches that are most similar to this thesis. They all focus on improving power consumption estimation in cloud computing environments by mean of calibration. Their approaches range from models based on crowd-sourced benchmarks to avoiding intrusion by completely relying on external power meters instead of hardware sensors and lastly to utilizing both external and internal.

Choochotkaew et al. [66] introduce a framework for enhancing energy consumption estimations in cloud environments through a container power model integrated with Kepler [24], an open-source tool for exporting energy-related metrics. This model addresses the challenge of variabilities in server power consumption due to dynamic background power, which traditional models struggle to accurately isolate. It employs a bidirectional approach that facilitates model training via crowd-sourced benchmarks for environments lacking direct online power measurements. The framework comprises three primary components: the extractor, isolator, and trainer modules, with the isolator module pivotal for estimating isolated power consumption using a new method that eliminates the need for profiling. The model's performance is quantified by its isolation goodness, which is the Pearson correlation between container resource usage and the isolated work-load power. The training process encompasses five steps: training a system power model, predicting background power, labeling power consumption, training the container power model, and enabling ongoing online model refinement. The model's accuracy is significantly improved by data contributions from various platforms, leading to a doubling of prediction accuracy in diverse environments and workloads compared to existing methodologies. This framework is technically integrated within this thesis as it is part of Kepler which is used in the approach. It is still included as it calibrates power-models to use for VMs in public cloud environments, where hardware sensors are not available for tenants. The main problem lies in the measurement source of the data the models are trained with, which are provided by the Kepler's offline approximation based on hardware sensors.

Guan et al. present WattScope, a system designed for non-intrusive monitoring of application-level power consumption within datacenters [34]. This system efficiently disaggregates power measurements at the server and rack levels, utilizing existing data from external meters embedded within power distribution units (PDUs). By avoiding direct interaction with hardware performance counters, WattScope remains non-intrusive. Central to its operation is a deep learning-based disaggregation technique adapted from building power systems, allowing precise separation of power readings into application-specific consumption data over time. WattScope incorporates a library of models fitted to divers application classes, facilitating integration with a cluster scheduler for appropriate model selection based on applications running on each server. The prototype is realized by adapting nilmtk-contrib, an open-source toolkit initially conceived for building energy disaggregation, to suit server and rack power environments. While WattScope does utilize a dedicated external socket-meter for calibration of its power models, it's operation requires selection of specific power models for each application, which is much more restrictive than the approach of this thesis.

The SMARTWATTS [33] system improves power consumption assessment of CPU and DRAM in computing environments with its system architecture shown in Figure 6. It provides detailed power esti-mations at the container level using Intel's RAPL and Linux perf with hardware performance counters. It employs automatic event selection to navigate counter limitations and uses Ridge regression for adaptive power modeling with real-time workloads. Static power is isolated by capturing HWPC data during idle states, refining it through median values. The implementation includes a client-side sensor, which collects HWPC and RAPL data at 2 Hz using Cgroups for monitoring, and a server-side power meter utilizing MongoDB for data storage and Scikit-Learn's Ridge regression for real-time and post-mortem power estimation analysis. This approach is likely the most similar approach to this thesis, as it uses a hardware sensor for approximation and a socket-meter for calibration of its power models. The main issue with this paper, however, is the lack of components included for its power monitoring approach and many missing features adding up to overall reduced accuracy compared to Kepler. SmartWatts calibration approach is also not applicable to multiple approximation sources as it is based on perf events[13] which is only applicable to the CPU.

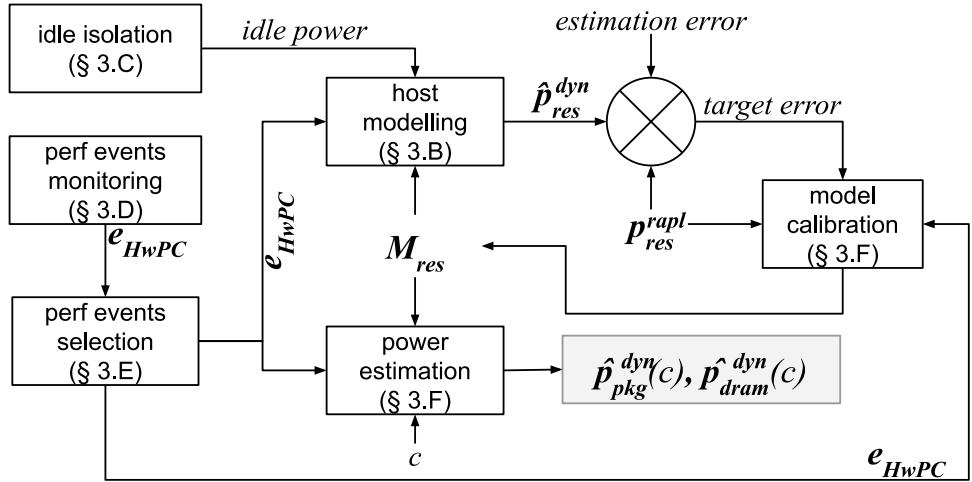

Figure 6: SmartWatts System Architecture [33]

---

[13]perf – Linux profiling with performance counters. https://perfwiki.github.io/main/

# Chapter 4

# Energy-Aware Testbed

This chapter presents the software testbed. The setup system facilitates the approximation of power consumption for individual applications, monitoring of their utilization of computing resources, and management of these resources to enforce power consumption limits. For calibrating the approximated process-level power consumptions, the total real-time system power is measured additionally. The combined application- and system-level visibility and control is integrated into a power system SiL co-simulation environment. Section 4.1 describes the architecture of the testbed and the interactions of its components. Section 4.2 then details the K8s implementation of the testbed, packaged a Helm chart and Section 4.3 lists the specific hardware and software requirements necessary to operate the testbed.

## 4.1 Architecture

To orchestrate computing nodes, this design uses a COP based implementation due to their popular status of uniformly managing resources of large server clusters. To streamline the setup process and enhance user accessibility, while leveraging standard industry tools, K8s is chosen as the COP. The architecture of the testbed is depicted in Figure 7. It is composed of four main components: applications, monitoring, simulation environment, and resource control, which interact to form a feedback loop.

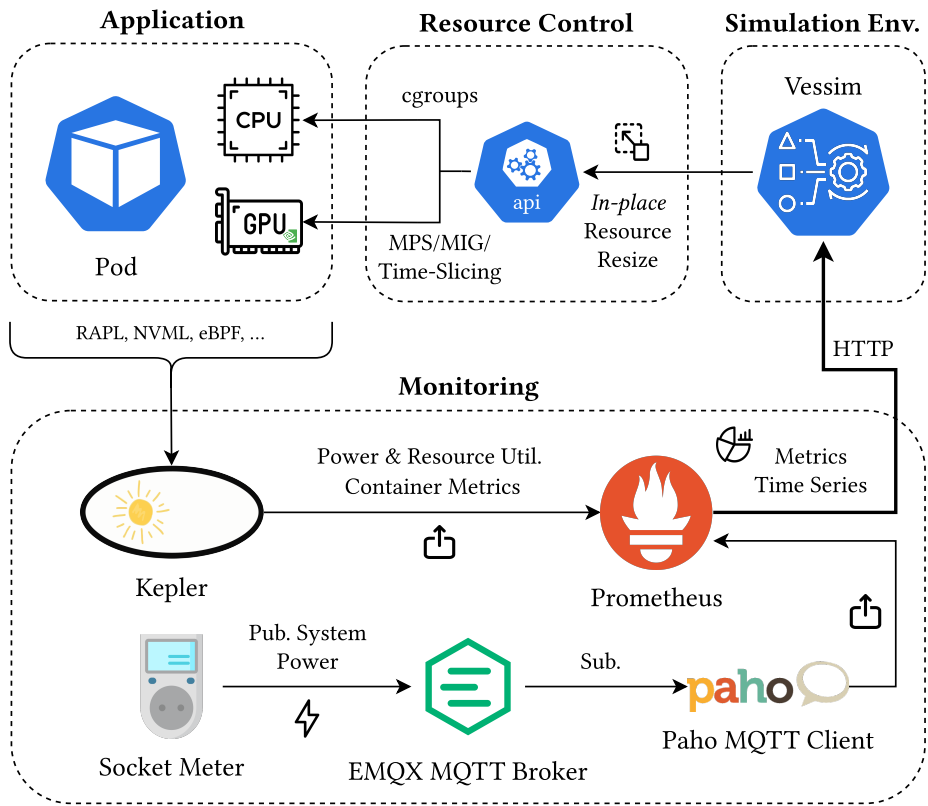

Figure 7: Testbed Architecture

### 4.1.1 Applications

Carbon-aware applications are deployed as K8s pods which can interact with the simulation environment's HTTP API server. They are able to monitor their power consumption, power system related metrics or other information provided through the configurable API such as current grid carbon intensity. Additionally, they can observe and manage their resource utilization, with particular emphasis on the GPU and CPU which are typically the main consumers of energy in computing systems.

### 4.1.2 Monitoring

The monitoring of these resources is facilitated by the Prometheus Exporter Kepler as described in Section 2.3.3. These power and re-source utilization metrics are committed to Prometheus as a cumulative monotonically increasing *counter*. In addition to individual pod power consumption approximations, a physical power meter device is installed between the computing node and the wall socket to which it is connected. The socket meter communicates via the MQTT protocol and publishes its real-time energy readings to the MQTT Broker *EMQX* [67]. While all other components of this testbed are deployed within a K8s cluster, the socket meter is a physical device and requires external access to the EMQX Service, managed by a K8s Ingress. A Paho-based [68] MQTT client then subscribes to the topic published by the meter, processes and finally exports the system power metric to Prometheus as a *gauge*, which represents a fluctuating single numerical value. Prometheus subsequently provides all these metrics as a time series, accessible through an HTTP API.

### 4.1.3  Simulation Environment

Vessim is specifically intended for the simulation of power systems, that interact with real or simulated computing systems [22]. Thus, it is integrated into this testbed for the execution of simulation scenarios involving power systems, including onsite renewable energy generation such as photovoltaic (PV) and wind energy, as well as energy storage like batteries.

As further detailed in Chapter 5, the metrics previously exported to Prometheus are now used to calibrate Vessim's microgrid simulation. Depicted in Figure 8, within Vessim's simulation, each application is represented as an Actor, denoted in red, which reports its consumed energy to the grid simulation and the Vessim Controllers, denoted in yellow. With the calibrated per container power consumption, Vessim provides visibility and control over the power system to these applications using an instance of the built-in `SiLController`, which deploys an HTTP API server that is accessible as the K8s Service `svc/vessim`. Both the power system and computing resources can be managed through the specified simulation scenario, for instance with a custom carbon-aware Controller, or defined API endpoints.

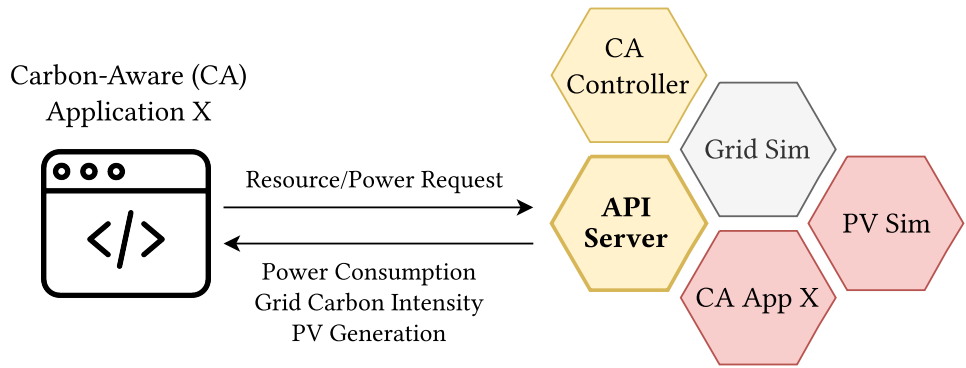

Figure 8: Application ↔ Vessim API

### 4.1.4 Resource Control

Computing resources can be managed with K8s' in-place vertical pod resizing of resource requests and limits for individual pods and their containers. For RAM and CPU resource caps, Linux cgroups are utilized, while for controlling GPU resources, currently only NVIDIA offers a sufficiently featured API for their GPUs. Depending on the specific use case and GPU model, either their Multi-Process Service (MPS), Multi-Instance GPU (MIG), or Time-Slicing methods can be used [69]. If the GPU is not shared between containers, power capping can also be applied with the NVML API instead [31]. Power capping on a process level, however, is facilitated by limiting the process's resource consumption. Many approaches to this issue already exist [70]–[72], [60], thus, this testbed does not address any specific implementations.

## 4.2 K8s Implementation

The software testbed is packaged as a Helm[14] chart to streamline deployment, configuration and dependency management, version control, and reusability. The implementation is outlined in Figure 9. Helm chart dependencies include the Kube Prometheus Stack[15], EMQX[16] and

---

[14]Helm – The package manager for Kubernetes. https://helm.sh/

[15]Kube Prometheus Stack Helm chart. https://github.com/prometheus-community/helm-charts/tree/main/charts/kube-prometheus-stack

[16]EMQX Operator Helm chart. https://github.com/emqx/emqx/tree/master/deploy/charts/emqx

NVIDIA GPU Operator[17], socket-meter and Kepler Exporter, any of which can be disabled to suit existing clusters. The Kube Prometheus Stack is a collection of K8s manifests, including the Prometheus Node Exporter which is required for the use of external Exporters that are used in this implementation, Grafana dashboards, and Prometheus rules and scripts to provide easy to operate end-to-end K8s cluster monitoring with Prometheus using the Prometheus Operator. The NVIDIA GPU Operator controls GPU resources and provides NVIDIA GPUs via CRDs. The socket-meter Exporter application as described above is packaged separately as it represents a stand-alone component. The Vessim simulation scenario is containerized and runs as a K8s Job, with a Service Account granted necessary permissions via a Cluster Role Binding. A K8s Service exposes the Vessim API within the desired scope and, finally, a Helm post-install hook (K8s Job) ensures that Prometheus successfully scrapes metrics from the Kepler and socket-meter Exporter.

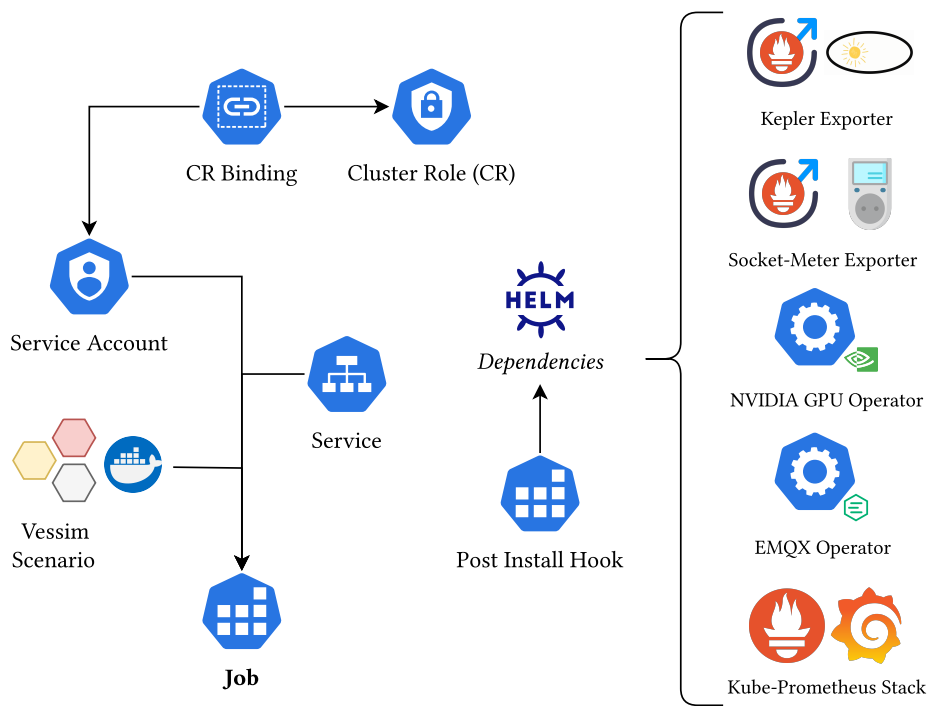

Figure 9: Testbed K8s Implementation

---

[17]NVIDIA GPU Operator Helm chart. https://github.com/NVIDIA/gpu-operator

## 4.3 Software and Hardware Requirements

To support the alpha feature for in-place vertical resource resizing, K8s must be version 1.27 or higher[18]. Given its alpha developments status, K8s has to be manually configured to enable the `InPlacePodVerticalScaling` feature gate. Installation on bare-metal is recommended for accessing hardware counters, although using a pass-through mechanism with sufficient permissions is also viable.

For energy readings, the CPU must support RAPL; AMD CPUs need to be at least Family 17th Processors[19] and Intel CPUs have to be from the Sandy Bridge architecture or newer [45]. With RAPL support, the CPU's architecture is x86. As outlined above, if a GPU is to be incorporated into the system, it should be an NVIDIA model. MPS requires the GPU's compatibility version to be 3.5 or higher [73] and MIG functionality is introduced with the Ampere generation [74], while the availability of Time-Slicing is unspecified by NVIDIA and is usually supported by most GPU models[20]. Finally, a physical power meter with a reasonably high sampling rate and precision is required, since these aspects directly affect the accuracy of the calibration procedure.

---

[18]K8s Blog Post. https://kubernetes.io/blog/2023/05/12/in-place-pod-resize-alpha/

[19]AMD Energy Kernel Driver. https://github.com/amd/amd_energy

[20]Time-Slicing GPUs in Kubernetes. https://docs.nvidia.com/datacenter/cloud-native/gpu-operator/latest/gpu-sharing.html

# Chapter 5

# Calibrating Vessim's Microgrid Simulation

The energy-aware testbed from the previous chapter provides approximations of power consumption and resource utilization for individual processes, and measures total system power consumption. This chapter demonstrates, in Section 5.1, how these metrics are utilized to calibrate simulations online for improved accuracy in representing per-process power consumption. Section 5.2 then describes the integration of these metrics into Vessim and explains the implementation procedure.

## 5.1 Calibration Procedure

To achieve accurate online calibration of per-process power consumption in Vessim's microgrid simulation, this approach draws on Kepler's Ratio Power Model, as outlined in Section 2.3.3. This modified method involves calibrating both the approximated idle and dynamic power consumption, recognizing that the socket power meter can only capture the *total* system power. The general calibration procedure for both idle and dynamic consumption involves multiplying an approximation factor $A \in (0, 1)$ that is the ratio of the approximated process to node power, provided by exported Kepler metrics, with the measured power component $M \in \mathbb{R}^+$ serving as the reference data:

$$p_{\mathrm{cal}} = A \cdot M$$

In system configurations where a component does not provide an idle consumption sensor, as Intel CPUs do with RAPL, Kepler approximates idle power through minimal resource utilization. The platform's idle power is directly obtained from the platform sensor. For instance, the host system in Chapter 6 uses ACPI as its platform sensor which reports a constant power consumption and is used as the idle value in platform measurements. Since system and K8s processes are operational during Kepler's idle measurements, it is necessary to record socket measurements concurrently to obtain $m^{\mathrm{idle}}$. The approximated idle process consumption can now be calibrated by multiplying the ratio of Kepler's approximated idle process power $p^{\mathrm{idle}}$ to that of the node's $n^{\mathrm{idle}}$ by $m^{\mathrm{idle}}$:

$$p_{\mathrm{cal}}^{\mathrm{idle}} = \underbrace{\frac{p^{\mathrm{idle}}}{n^{\mathrm{idle}}}}_{A} \cdot \underbrace{m^{\mathrm{idle}}}_{M}$$

Similar can be applied to calibrate the dynamic process consumption. However, as further examined in Section 6.3.2, Kepler reports that approximated dynamic system processes' power $s^{\mathrm{dyn}}$ scales proportionally to the dynamic workload process power $p^{\mathrm{dyn}}$. This behavior needs to be reflected in the calibration procedure by attributing the portion of the responsible system's dynamic power consumption to the process's. Determining the exact attributable consumption to workload processes is complex due to unknown involvement of other system processes. However, it can be reasonably assumed that power usage is proportionally shared among workload processes based on their relative activity levels. The non-attributed power of other workload processes is given by the nodes' consumption subtracted from the sum of the observed process and system processes' power and therefore yields:

$$A = \left( p^{\text{dyn}} + \frac{p^{\text{dyn}}}{p^{\text{dyn}} + n^{\text{dyn}} - (p^{\text{dyn}} + s^{\text{dyn}})} \cdot s^{\text{dyn}} \right) \cdot \left( n^{\text{dyn}} \right)^{-1}$$

$$= \frac{n^{\text{dyn}} \cdot p^{\text{dyn}}}{n^{\text{dyn}} - s^{\text{dyn}}} \cdot \left( n^{\text{dyn}} \right)^{-1}$$

$$= \frac{p^{\text{dyn}}}{n^{\text{dyn}} - s^{\text{dyn}}}$$

Now the calibration factor needs to be multiplied with the difference of the socket-meter measured total system power consumption $m$ and the measured idle system power consumption $m^{\text{idle}}$ which finally yields:

$$p^{\text{dyn}}_{\text{cal}} = A \cdot \underbrace{\left( m - m^{\text{idle}} \right)}_{M}$$

## 5.2 Vessim Implementation

With this calibration procedure, this section highlights the integration process by which Vessim utilizes signals and actors to accurately simulate the idle and dynamic power consumption of running processes.

### 5.2.1 Signals

Vessim's actors can determine their behavior as either producers or consumers through the use of signals as interfaces. These signals typically represent historical traces that are advanced according to predefined simulation steps, returning data points at specific times. In real-time SiL simulations, like this approach is based on, these signals can also be used to provide data points corresponding to the *current* time. Metrics can be collected via Prometheus and are accessible through HTTP as depicted in Figure 7. This behavior is implemented by the `PrometheusSignal` class, which is a subclass of Vessim's `CollectorSignal`, and is designed to execute PromQL queries to obtain a desired metric. This class operates by updating the signal's returning value at set intervals within a separate thread, ensuring that Vessim's main thread remains unblocked to avoid temporal discrep-

```
 1  class COLLECTORSIGNAL (interval) ←   class PROMSIGNAL (query, interval)
 2  │ v = 0.0                               │ super (interval)
 3  │ interval = 1 if not set               │ function COLLECT ()
 4  │ start COLLECTLOOP as a thread         │ │ return HTTPPROM (query)
 5  │ function NOW ()
 6  │ │ return v
 7  │ abstract function COLLECT ()
 8  │ function COLLECTLOOP ()
 9  │ │ loop
10  │ │ │ v = COLLECT ()
11  │ │ │ sleep for interval
```

Algorithm 1: Signal Classes Implementation

ancy with expected real-time progression. The details of this implementation are illustrated in Algorithm 1.

### 5.2.2 Actors

In this approach, actors are designed to manage the consumption associated with individual processes or groups of processes. These actors utilize `PrometheusSignals` as the basis for their consumption decisions and for implementing calibration procedures. In order to properly calibrate consumption, each necessary source is accessed through a `PrometheusSignal`. However, querying the Prometheus server for each data point during each interval introduces overhead, so only the necessary procedures are implemented to manage this resource efficiently. For instance, in energy-aware scenarios, idle power consumption might not be relevant as it is unaffected by process actions during runtime. Instead of monitoring every process or pod within a namespace, which could be unnecessarily exhaustive, it might be more efficient to monitor the consumption of the entire application group by focusing on the namespace's collective usage. This method is illustrated in Algorithm 2, which shows the calibration process for all processes within a shared namespace, executed through the `NamespaceActor` class implementation, a subclass of Vessim's `ActorBase` class. As previously mentioned in Section 2.3.3, every second, a moving average within the interval of 2 seconds is applied to the PromQL query to slightly mitigate the effects of Kepler's fluctuating metrics.

```
 1  class NamespaceActor (namespace)
 2      query = "sum(rate(kepler_container_platform_joules_total{"
 3      interval = "}[2s]))"
 4      sig_m = PromSignal ("socker_meter")
 5      sig_n^dyn = PromSignal (query + "mode='dynamic'" + interval)
 6      sig_p^dyn = PromSignal (query +
 7          | "container_namespace=namespace, mode='dynamic'" + interval
 8      )
 9      sig_s^dyn = PromSignal (query +
10          | "container_namespace='system', mode='dynamic'" + interval
11      )
12      m^idle = sig_m → NOW ()
13      function P ()
14          m = sig_m → NOW ()
15          n^dyn = sig_n^dyn → NOW ()
16          p^dyn = sig_p^dyn → NOW ()
17          s^dyn = sig_s^dyn → NOW ()
18          return  (p^dyn / (n^dyn − s^dyn)) · (m − m^idle)
```

Algorithm 2: Actor Calibration Implementation

# Chapter 6

# Evaluation

This chapter examines the testbed's ability to calibrate energy consumption across diverse workloads and begins by detailing the benchmark setup, which is designed to assess the behavior of the hardware components under different utilization profiles. Using a robust host system configuration featuring an Intel Xeon CPU and NVIDIA Quadro GPUs, the experiments simulate real-world cloud service demands by deploying a Kubernetes-centric architecture with CPU and GPU intensive workloads based on the frameworks DeathStarBench and Ray. These are supplemented by baseline monotone stress tests using the tool stress-ng. The results of these benchmarks are examined with a focus on the accuracy of total system power approximation and the validation against measurements from the socket power meter. The approximations are compared with actual measurements using regression analysis to assess their reliability. Additionally per-process power calibration is explored, which is especially relevant for multi-process environments typical in cloud computing. The chapter concludes by reviewing constraints that interfere with the calibration procedure and discussing the testbed's alignment with initial requirements.

## 6.1 Benchmark Setup

The host system is powered by an Intel Xeon "Cascade Lake" Silver 4208 CPU, running with 8 cores and 16 threads at 2.10 GHz with an LGA 3647 socket. It is equipped with 6 × 16 GB DDR4 RAM modules, limited to 2400 MHz by the CPU. Storage is provided by 2 × 1TB Micron 2200 SSDs, using M.2 NVMe PCIe 3.0 interfaces. Graphics are handled by 2 × NVIDIA Quadro RTX 5000 GPUs with 16 GB GDDR6 each, based on the Turing architecture. Networking includes 2 × 10 GBit Ethernet NICs, namely the Intel X550. The operating system is Ubuntu 24.04.1 LTS. The server is connected to a *GUDE Expert Power Control 8045 PDU*, a 12-fold switched and outlet-metered unit [75]. This PDU includes the *socket-meter* as referred to in previous chapters, which is connected to the same local network as the server for publishing measured power consumption with minimal latency. GUDE reports electrical measurement inaccuracies and sampling rates of < 1% for voltage (V) at 0.01s, < 1.5% for current (I) at 0.001s, and < 1% for phase shift ($\varphi = \arg(V) - \arg(I)$) at 0.1s. All power consumption measurements in this evaluation are reported as active power ($P$), measured in watts (W), as it is the *actual* power consumed by the server [76], given by the formula:

$$P = V \cdot I \cdot \cos(\varphi)$$

This benchmark design, as illustrated in Figure 10, mandates a K8s-centric architecture, where each benchmark is packaged as a Helm chart and executed within a dedicated namespace, "bench" within this evaluation. This approach accommodates common cloud service benchmarking patterns, featuring load generators designed to stress the System Under Test (SUT) [77]. Within the benchmarking namespace, resources are continuously monitored as detailed in Chapter 4, allowing real-time metrics provision to a Vessim calibration actor located in the "energy-aware" namespace to refine the approximated power consumption as described in Chapter 5. Unlike a push model, which necessitates privileged host access, a pull model for results collection by the host system is adopted to maintain security by avoiding granting unnecessary access rights to the Vessim pod. The deployment and management of benchmarks are streamlined via the Helm CLI, facilitating easy initiation, termination, and configuration during the simulation through the `HelmBenchmarkController` class, implemented as shown in Algorithm 3. This abstract class is a custom controller derived from Vessim's `Controller` abstract class, which oversees the lifecycle of benchmarks and synchronizes data with Vessim's Monitor. Each specific benchmark extends this custom controller class, implementing the `updateValues()` method to adjust the configurations for subsequent runs. While resource control via the K8s API is technically feasible, it is inadvisable in the context of a benchmark with a GPU intensive workload to adjust power consumption directly. Restricting the performance of a CPU on a process level works by limiting the CPU time utilized by that process [78]. The same process is vastly more inefficient for GPUs as they are designed for parallel processing, thus their performance degrade under such constrained resource conditions [79]. Instead, optimization should focus on workload modulation, such as reducing batch sizes.

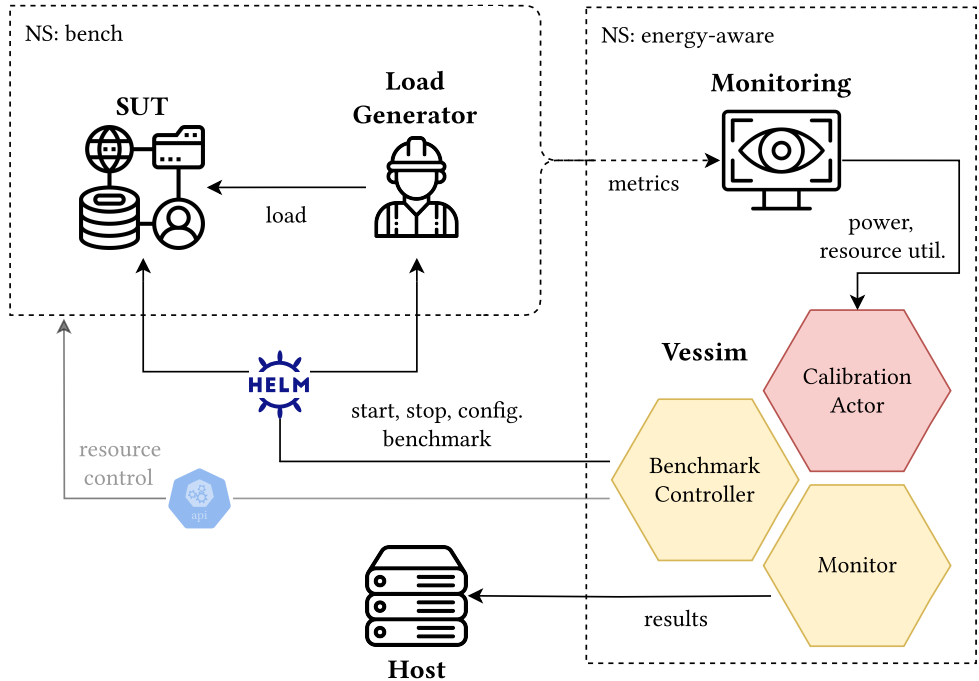

Figure 10: Benchmark Architecture

## 6.2 Workloads

This evaluation is based on both CPU and GPU intensive workloads. To ensure fairness between the different benchmarks, the control of

```
1  abstract class HELMBENCHMARKCONTROLLER (runtime)
2      startTime = runtime + 1
3      abstract function UPDATEVALUES ()
4      function STEP (time)
5          if time - startTime <= runtime then
6              return
7          if benchmark is running then
8              STOPBENCHMARK ()
9          UPDATEVALUES ()
10         STARTBENCHMARK ()
11         startTime = time
```

Algorithm 3: Helm Benchmark Controller Implementation

power consumption for *all* workloads is based on modulation, opposed to resource limitation. Different types of workloads stress system components differently, while similar resource utilization can still result in varying power consumption due to distinct operations performed on the underlying hardware [32]. Since the approximation of power consumption in this approach is primarily based on resource utilization of each component, a diverse set of workloads with a heterogenous hardware utilization profile is necessary to evaluate the system under different conditions. Additionally, workloads should allow to accommodate the scalability of the testbed. The workload matrix is illustrated in Table 1 with the frameworks used in this evaluation being the *DeathStarBench* (DSB) benchmarking suite, the distributed machine learning compute engine *Ray* and the tool *stress-ng*.

### 6.2.1 DeathStarBench

Large-scale datacenters are often hosting popular cloud services that demand stringent performance, latency, and availability requirements. This has caused monolithic application design to shift to microservices, where applications are divided into numerous single-purpose, loosely-coupled components. Microservices introduce challenges across the cloud system stack, from hardware to application design, which the DeathStarBench suite helps to examine by quantifying the implications of these microservices on hardware, networking, operating systems, cluster management, and programming frameworks [80]. DSB includes six services, with three having reached stable release status, and two of them are working with K8s. For the purpose of this evaluation, these two specific services will be utilized: the *media* and *hotel reservation service.*

| Target Resource | Bench Suite | Base Metric | Workload |
|---|---|---|---|
| CPU | DSB | RPS | Hotel Reservation |
| | | | Media |
| | stress-ng | Threads | *CPU* |
| | | | *VM* |
| | | | *IO* |
| GPU | Ray | Batch Size | Fashion-MNIST |
| | | | Fashion-MNIST |
| | | | CoLA |

Table 1: Workload Matrix

The media service application provides a platform for users to browse, review, rate, rent, and stream movies. The system architecture, shown in Figure 20, Appendix A, involves users sending HTTP requests through a load balancer and web servers, both implemented with nginx[21], which subsequently utilize the PHP-FPM[22] module to communicate with various microservices via Apache Thrift[23] RPCs. The service facilitates movie rentals through a payment authentication module that ensures sufficient user funds, followed by video streaming via the nginx-hls module for HTTP live streaming. Movie files are

[21]nginx – HTTP web server, reverse proxy, content cache, load balancer, TCP/UDP proxy server, and mail proxy server. https://nginx.org/en/

[22]PHP-FPM – simple and robust FastCGI Process Manager. https://php-fpm.org/

[23]Apache Thrift – framework for scalable cross-language services development. https://thrift.apache.org/

stored on a Network File System (NFS) to minimize access latency and complexity, while reviews are maintained in memcached[24] and MongoDB[25] databases. Movie information is stored in a sharded and replicated MySQL[26] database. Finally, the application includes recommendation systems for movies and advertisements, supported by auxiliary services for maintenance and service discovery.

The hotel reservation application is a smaller solution in Golang designed to simulate the process of reserving a hotel room, comprised of 8 services [81]. It enables users to browse through hotel information and complete reservations, providing functionalities like filtering based on ratings, price, location, and availability. The system also offers personalized hotel recommendations to its users. The client requests initially interact with a front-end web server, which routes them to appropriate service layers for hotel searching, reservation completion, and recommendation generation. The architecture supports geolocation-based hotel searches and uses memcached for in-memory data caching and MongoDB for longer-term data storage. Communication between service tiers employs the gRPC protocol.

wrk2[27] is the load generator component of the DSB suite and based on the open-source benchmarking tool, wrk[28]. It has been modified to act as an open-loop load generator, i.e. it can send out requests on a fixed schedule, regardless of the completion state of previous requests. By running on a client system, wrk2 generates multiple HTTP requests at once to a DSB application. Having the load generator and the SUT run on the same machine is usually considered an anti-pattern in benchmarking, as it can lead to skewed results due to resource contention [82]. Because of this, a separate machine with a 32 threads AMD EPYC 7282 16-core CPU, is introduced to the K8s cluster to host the wrk2 application. The load generator for both workloads has been configured to incrementally escalate the number of requests per second (RPS) from an initial rate of 250 to a maximum of 2000 over the course of 8 benchmarking iterations, each lasting 10 minutes. The requests are evenly distributed across 32 concurrent connections directed towards the SUT. This distribution operates across all threads of the generator's host system, with each thread maintaining a single connection as recommended by the DSB development team[29]. Both workloads are tested with mixed workload types provided by the DSB suite.

---

[24]memcached – distributed memory object caching system. https://memcached.org/

[25]MongoDB – document database. https://www.mongodb.com/

[26]MySQL – relational database. https://www.mysql.com/de/

[27]wrk2 – HTTP benchmarking tool based on wrk. https://github.com/giltene/wrk2

[28]wrk – HTTP benchmarking tool. https://github.com/wg/wrk

### 6.2.2 Ray

With machine learning workloads becoming more compute-intensive, single-node development environments often fail to adequately scale to meet demands. Ray is an open-source framework that provides a robust compute layer that enables parallel processing and efficient scaling of AI and Python applications [83]. Different ML batch workloads are examined using distributed PyTorch training on Fashion-MNIST for CPU and GPU and CoLA dataset for GPU.

The MNIST dataset [84], remains a popular tool in deep learning due to its simplicity and ease of use, despite the availability of more complex datasets like CIFAR-10 [85] and ImageNet [86]. Recognizing the need for a similarly accessible yet more challenging benchmark, the Fashion-MNIST dataset was developed. Fashion-MNIST retains the original MNIST format, 10 classes of 70,000 grayscale images at 28 × 28 resolution, but incorporates images of fashion items from Zalando, offering a more complex classification task compared to the original digit dataset [87]. The images undergo a conversion process to standardize them, including resizing, sharpening, and converting to grayscale, before being divided into training and test sets.

The Corpus of Linguistic Acceptability (CoLA) is a curated dataset of sentences from linguistic literature, each labeled for grammatical acceptability [88]. CoLA covers a wide range of syntactic and semantic phenomena, sourced from textbooks and research publications.

---

[29]Yu Gan's Comment on DSB GitHub Issue #3. https://github.com/delimitrou/DeathStarBench/issues/3#issuecomment-493701839

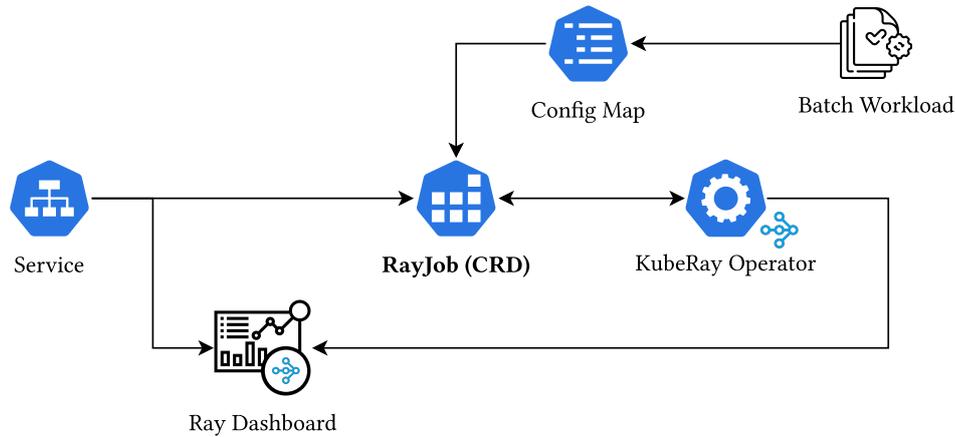

Figure 11: Ray Benchmark Architecture

Sentences are adjusted to exclude out-of-vocabulary words, and the dataset is divided into in-domain and out-of-domain sets to monitor overfitting. Annotations for various linguistic features ensure balanced representation. Human evaluations show an 86.1% agreement with CoLA labels, highlighting its reliability despite some dialectical variations.

In the context of evaluating batch workloads using Ray, the general benchmark design from Figure 10 which largely applies to continuous service-based loads, may not offer direct applicability to the batch job environment. Unlike service workloads, batch jobs do not necessitate a load generator since they execute pre-defined tasks at specified times. Figure 11 presents the benchmark architecture specific to Ray, where batch workloads are designated through their source code and referenced in a ConfigMap. This ConfigMap is then mounted onto a RayJob Custom Resource Definition (CRD), alongside a Service which facilitates dashboard access for users. The Ray Operator observes and schedules these RayJob CRDs, orchestrating workload execution on the cluster while managing the lifecycle of the resources involved. Controlled variation in resource consumption is achieved through adjustments in batch sizes [89]. Each benchmark iteration persists for ten minutes, evaluating batch sizes of 1, 4, 8, 16, and 32. While the CoLA dataset is trained as a GPU-based workloads, Fashion-MNIST is trained for both CPU and GPU as the dataset is small enough.

### 6.2.3 Baseline Stress Tests

The tool stress-ng[30] is utilized to establish a baseline for hardware utilization profiles by methodically stressing specific CPU resources. The tool is configured to operate with a consistent number of threads, each executing designated operations on the CPU. Three particular execution options are chosen for this assessment: *CPU, IO,* and *VM*. For *CPU* stress testing, methods are sequentially worked such as mathematical and algorithmic tasks including the Ackermann function, matrix multiplication, and fast Fourier transforms. The *IO* operations comprise a combination of read/write activities – both sequential and random – and included complex tasks like forced synchronization and cache management, with multiple child processes interacting with a shared file. The *VM* operations incorporate memory testing methods using techniques like bit flipping and gray codes. Tests are conducted by incrementally increasing the number of threads from 1 to 16 with 10 minutes of execution time on the host system to monitor the differing levels of power consumption. The CoLA dataset is only benchmarked as a GPU-based workloads due to its computational complexity. The Fashion-MNIST dataset is compact enough to be efficiently trained on both CPU and GPU systems.

## 6.3 Results

The evaluation of benchmark results is conducted in two parts. First, the total power approximation for the whole system is validated. This involves assessing the accuracy of approximations provided by the Kepler framework using various hardware counters and sensors in the host system. This provides insights into any patterns between approximated and measured consumption. However, there is currently no definitive method to determine the power consumption of individual processes, making it difficult to validate per-process power consumption estimates as they are based on resource consumption rates. With this limitation, the second part involves the calibration of these per-process power approximations to improve the understanding of process-specific power usage over existing methods. This calibration method aims to address observed issues in the Kepler framework and is evaluated by demonstrating typical use cases and comparing differences to non-calibrated workloads.

---

[30]stress-ng – Stress test tool. https://github.com/ColinIanKing/stress-ng

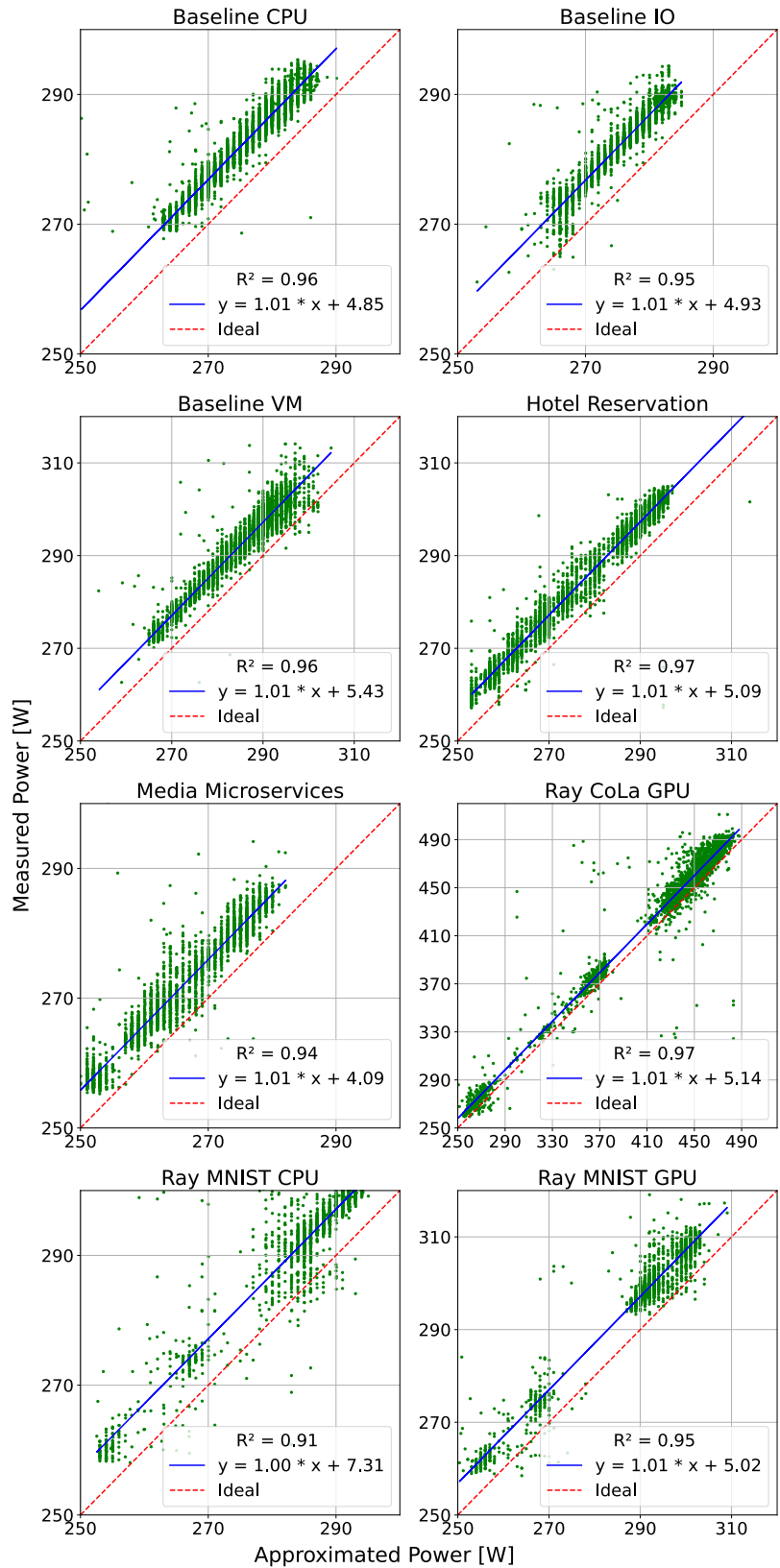

Figure 12: Approximations vs Measurements with Regression Line $y$

### 6.3.1 Validation of Total System Power Approximation

The first series of benchmarks are conducted using the socket power meter as the definitive measure of the actual power consumption of the system. The initial set of benchmarks involves plotting the approximated power consumption on the x-axis against the measured power consumption on the y-axis, which is depicted in Figure 12 with the units being Watts. To determine the precision of the approximation, a regression analysis is performed. A regression line is fitted to the observed data points and subsequently compared with the ideal line represented by the graph $y = x$. The observed data patterns suggest a linear relationship between the measured consumption, treated as the dependent variable, and the approximated consumption, functioning as the regressor. Consequently, a linear regression model is used to fit the data points.

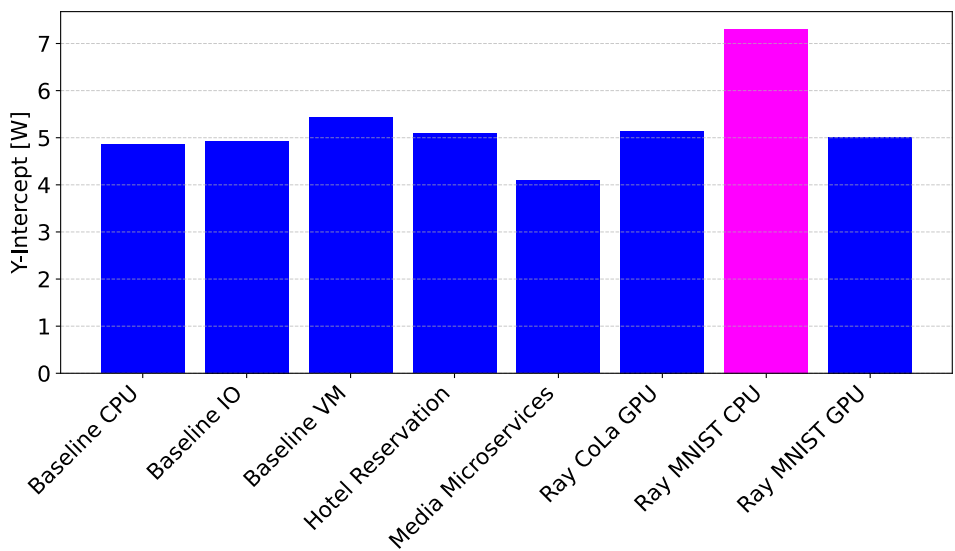

Figure 13: Y-Intercept of Regression Line

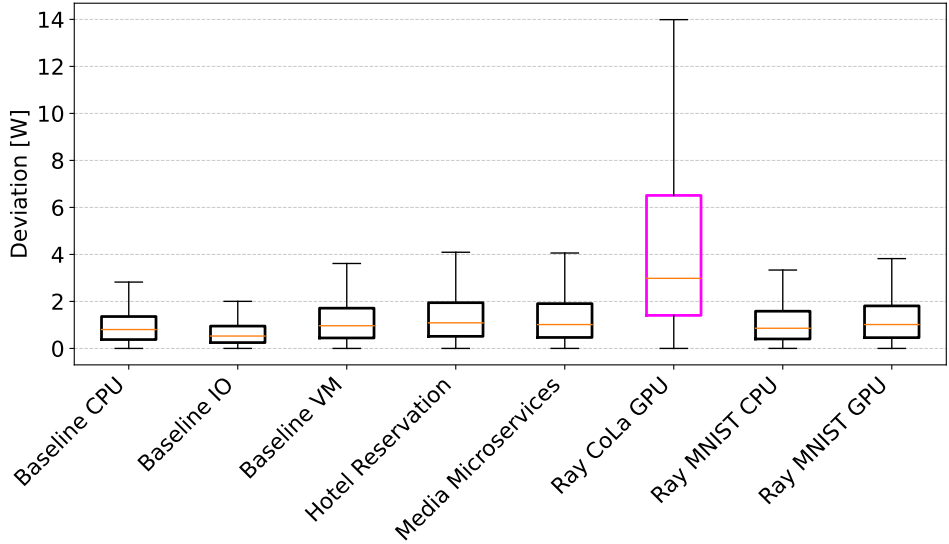

Figure 14: Deviation from Regression Line

The results indicate a strong relationship between Kepler's approximations and the actual measured power consumption, as evidenced by the average regression coefficient of approximately 1.01 across all test cases. This suggests that the approximations serve as reliable predictors of power consumption. The y-intercept of the regression lines is employed to examine the deviation from the measured values, with an average y-intercept depicted in Figure 13 of ~5.23 W. Figure 12 also shows the $R^2$ values for each workload, with an average of 0.95, which indicates a good fit of the data points to the regression line. Notably, the Fashion-MNIST benchmark deviates from both the average y-intercept and $R^2$, exhibiting a y-intercept of 7.31 W and an $R^2$ of 0.91. Figure 14 illustrates the variances from the regression lines through box plots, showing that the workloads have a median deviation of ~1.16 W on average from the regression line. However, the Ray CoLA benchmark stands with a considerably higher median deviation of 2.98 W and a maximum deviation reaching 14 W, in contrast to other benchmarks that demonstrate maximum deviations generally between 2-4 W.

This observed relationship reveals a promising potential for simplified calibration processes. The generally close y-intercepts is likely indicative of the system's idle power consumption and as it does scale with higher power consumption as exhibited the CoLa workload. Paired with the near-ideal regression coefficient, $R^2$ values and the minimal deviations from the regression line, a one-time initial calibration might suffice for accurately predicting power consumption across various workloads. However, certain machine learning workloads present as outliers, showing greater deviations that exceed typical boundaries. While these deviations, such as those observed in the Ray CoLA benchmark, indicate a necessity for refined adjustments, they can potentially be addressed through a general calibration process applicable across all workloads with a minimal error margin. For enhanced precision, these outliers may benefit from a more targeted, workload-aware calibration.

### 6.3.2  Per-Process Power Calibration

As mentioned in Chapter 5, the dynamic system processes exhibit a proportional scaling with the workload processes of a GPU, as presented in Figure 15. This figure shows a 5-minute analysis with the CoLa benchmark as the GPU workload and the DSB hotel reservation service for the CPU workload, demonstrating fluctuations in node power consumption resulting from varying workload stress levels within the selected window. Specifically, in the context of the GPU workload, an increase in batch size from 1 to 4 triggers a corresponding increase in power consumption. Similarly, the CPU workload displayed elevated power use in response to an escalation from 250 to 500 RPS generated by the load generator. Notably, for the GPU workload, the system processes, identified by the orange `sys_dyn` plot, evidently scale in direct proportion to the workload (or namespace) processes, indicated by the blue plot `ns_dyn`, as observable during the intervals from minutes 9-10 and 12-14. Conversely, such proportional scaling was not consistently evidenced in the CPU workload, where the relationship was apparent only during specific workload process surges. The observed proportional scaling between GPU workload processes and system processes require further investigation to explain the underlying mechanisms.

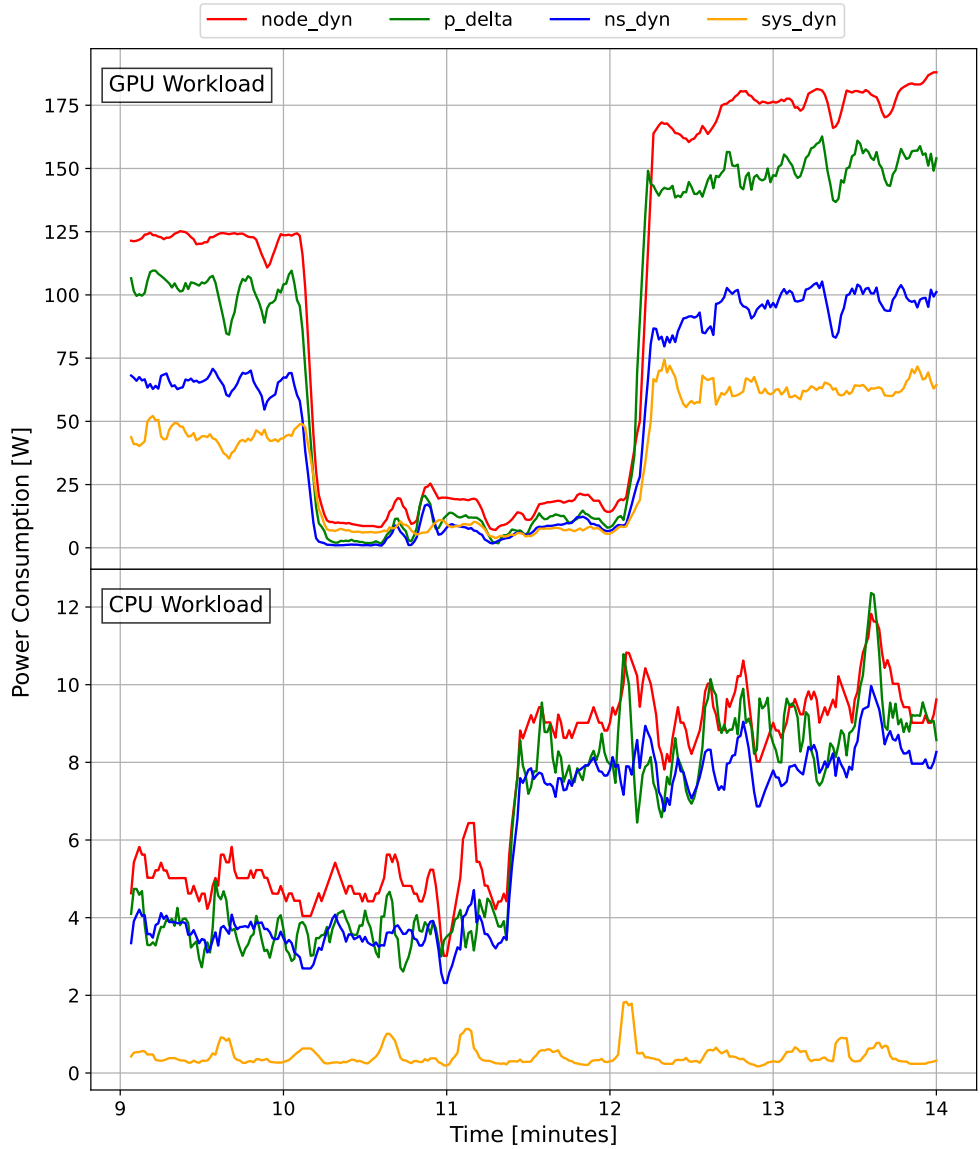

Figure 15: System Processes' Power

This phenomenon can potentially be attributed to the special nature of GPUs and their complex interactions with system components, as opposed to the more tightly integrated CPUs and memory modules. The GPU-specific kernel activities, such as scheduling, memory management, and data transfer, might elevate system process activity levels [90]. Additionally, the mandatory incorporation of a manufacturer-specific GPU operator within K8s clusters could introduce further overhead, impacting power consumption dynamics. For the current state of art, the attributed power consumption of these system processes is proportionally distributed among all active workload processes, weighted by their individual power consumption metrics, for calibration purposes. This procedure is depicted by the green plot `p_delta`, though the weighting effect can not be measured here due to only a single workload process being operational. Figure 16 illustrates the considerable impact, the calibration has on the workload by comparing the proportions of Kepler's approximation and the calibration offset, together summing up to the total calibrated energy consumed during the benchmark, as the offset constitutes to about $\frac{1}{3}$ of this energy.

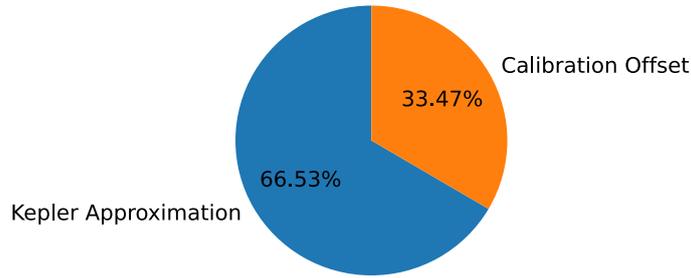

Figure 16: GPU Workload Calibration Offset

In practical applications, understanding the power consumption of workloads is crucial, particularly when multiple processes are executed simultaneously, as is typical in cloud computing environments. Measuring the power consumption of a single workload can be straightforwardly achieved using the socket-meter; however, determining the power utilization of individual processes among several concurrent workloads is significantly more challenging. A microservice architecture exemplifies this scenario, wherein numerous applications operate concurrently within the same system. Figure 17 depicts a 3-minute window during which the DSB hotel reservation service is executed. During this period, the load generated is increased from 1000 to 1250 RPS, and the power consumption of each microservice is calibrated and shown. Through the visualization, it can now be observed that, despite the load generator imposing a mixed workload where 60% of requests targeted the search service, 39% the recommend service, and a mere 0.5% each to the user and reservation services, the reservation service exhibited the highest power consumption among all. Additionally, certain services, such as MongoDB and memcached, which exhibit minimal power consumption, were easily identifiable. In the representation, any service consistently using less than 2 W throughout the timeframe is indicated in grey. Figure 18 illustrates the aggregate calibrated energy consumption of all microservices over the entire 80-minute benchmark period, resulting in a total usage of 32.33 Wh, about 3.5% more than the consumption approximated by Kepler. The data reveals that the reservation service accounted for a significant 42.6% of total consumption e.g. including 0.47 Wh of calibrated energy, followed by the frontend at 15.6%, with the rate and search services using 12.7% and 11.1% respectively. Meanwhile, other services, such as the databases, with an aggregate consumption of less than 1 Wh each, are collectively categorized as *other* services.

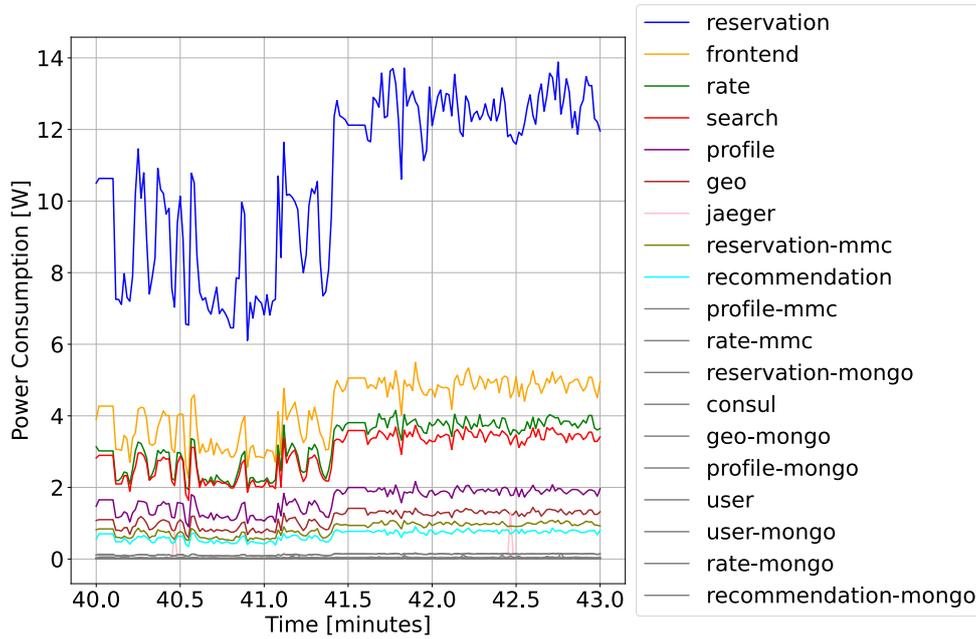

Figure 17: Per-Process Power Time Series

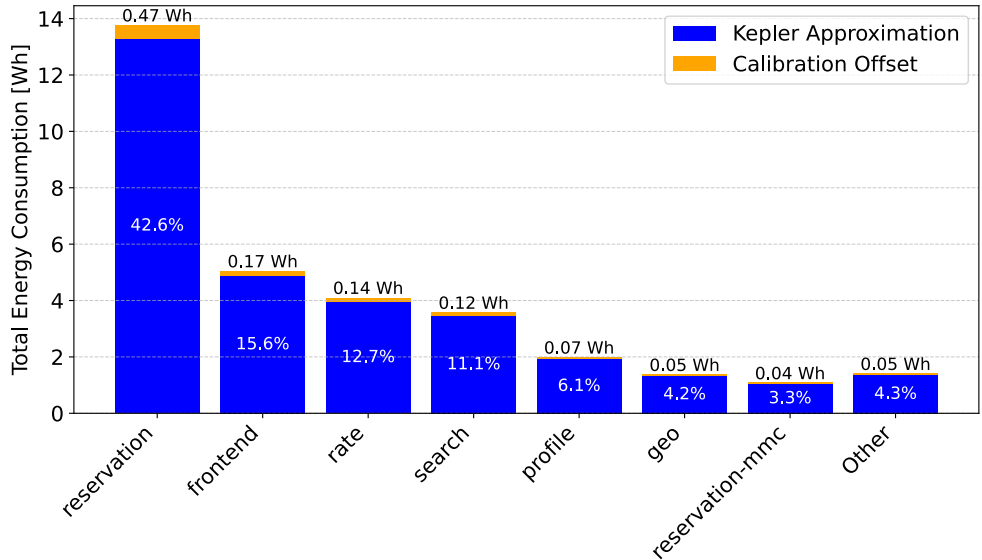

Figure 18: Per-Process Total Energy Consumption

### 6.3.3 Restrictions

Several constraints which interfere with the accuracy of validation and approximation procedures have been identified. Notably, current GPU drivers and integrations in K8s and computing systems in general, are suboptimal with frequent issues reported by users, with a heavy dependency on the specific GPU manufacturer. Although anticipated advancements in GPU technology – driven by a high demand in AI-related needs – hold promise for resolving these challenges, the source of existing GPU power attribution issues remains uncertain, whether they stem from the Kepler framework or the broader GPU-system integration. Interestingly, GPU hardware sensor readings demonstrate a high accuracy, as observable in Figure 12 as the regression line would otherwise show significant differences between GPU and CPU workloads. Furthermore, determining the system's idle power consumption is complex due to user-dependent definitions of *idle.* In this evaluation, idle power consumption is assumed to be measured in the absence of active workload processes, resulting in an average value of 260 W while only the system and K8s cluster functions are active. However,

Kepler does not clearly define its methodology for measuring idle power, contributing to an ambiguous understanding. Additionally, the employed socket meter exhibits a combined maximum error rate of 3.5% when measuring the active power of the host, necessitating careful consideration in result interpretation.

## 6.4 Conformity with Requirements

While the ability of the energy-aware testbed to calibrate power consumption has been demonstrated above, the testbed's conformity with the requirements as outlined in Section 1.1 remains to be evaluated.

Requirement (1), focusing on monitoring power generation and consumption (1.1), the testbed mandates the use of a socket power meter for tracking server power consumption. The prototype employs the GUDE Expert Power Control 8045 PDU, notable for its minimal measurement inaccuracies and high sampling rates of < 0.1s. The testbed also supports monitoring power consumption of individual processes via the open-source K8s native framework, Kepler. In terms of monitoring carbon intensity in real-time (1.2), Vessim facilitates the tracking of carbon intensity using APIs such as WattTime and ElectricityMaps. These tools aid in obtaining carbon intensity of the electrical grid, which can be aligned with server or process power consumption to assess emissions. Regulation of power usage based on this carbon intensity (1.3) is achieved through *In-Place Resource Resize*, as further elaborated in (3), allowing the testbed to adjust resource utilization and, in turn, power usage according to carbon intensity insights. Vessim also enables simulation of batteries and renewable energy sources through its built-in models or customized additions, to leverage metrics these metrics to guide scenario behavior within the simulation environment.

Requirement (2) focuses on the orchestration of applications using the COP Kubernetes (K8s) system. This allows for the management of applications with minimal adjustments, requiring only their containerization (2.1). As a result, applications can be orchestrated without altering their core implementations. There is no need for applica-tions to measure or report their power usage. The Kepler framework provides non-intrusive monitoring of power consumption, to simplify the process for application developers and yielding streamlined operations within the testbed (2.2).

Requirement (3) applies to dynamic vertical scalability, allowing for the adjustment of resources allocated to a running pod without requiring a restart. This is achieved through K8s' new *In-Place Resource Resize* feature, which is accessible through the K8s API during simulation scenarios. The flexibility of this is further shown in combination with the Vessim API Controller, which allows applications to request a change in resource allocation as needed.

Requirement (4) of the testbed emphasizes economical viability paired with accessibility and simplicity for users. Under inexpensiveness (4.1), the testbed is compatible with most x86-based systems, including less powerful hardware, as it is not overly resource-demanding. Detailed system requirements are provided in Section 4.3. Although the professional-grade socket power meter used in the prototype, priced at approximately 1000€, is costly and designed for high accuracy and sampling rates across multiple servers, the testbed does not necessitate other expensive external hardware, unlike prior Ecovisor implementations from Souza et al. which require cost-intensive components like a solar array emulator for 10.000$. The impact of lower sampling rates and accuracies is acknowledged as an area for future investigation Chapter 7. Meanwhile, regarding simplicity in setup and maintenance (4.2), the testbed benefits from a straightforward installation process. All components are containerized and managed via K8s, with deployment facilitated through a Helm chart that requires minimal manual configuration. From the use of K8s the testbed also inherits features such as self-healing, scalability, and streamlined operations.

# Chapter 7

# Conclusion

This thesis introduced an energy-aware software testbed that employs the Software-in-the-Loop (SiL) co-simulation framework, Vessim. This testbed facilitates the assessment of energy- and carbon-aware computing systems and strategies, to incorporate renewable energy sources such as solar and wind via Vessim's SiL capabilities or through its integrated or external simulators. Within the simulation environment, both simulated and real computing nodes can be represented to allow for real-time measurement and control of their power consumption. Approximations of power usage at the system- or application-level are realized through the Kubernetes(K8s)-native, open-source framework Kepler, complemented by a socket power meter that provides a definitive measurement for the entire node's energy consumption. These metrics are utilized to calibrate the estimated power consumption of individual running processes within Vessim's microgrid simulation scenario in real-time. The orchestration and control of computing nodes or specific applications are managed through the container orchestration platform, K8s.

The testbed distinguishes itself from existing energy-aware systems by offering a cost-effective yet realistic solution that combines the precision of physical computing systems with the economically feasible simulations of costly microgrids. It adequately incorporates Ecovisor capabilities as defined by Souza et al. [13] while inheriting K8s' inherent scalability, which allows for simple expansion without adding complexity or incurring exponentially increasing costs. Additionally, the testbed employs an online-calibration methodology, which yields a more accurate depiction of power consumption for computing nodes and individual processes within the simulation.

The testbed was evaluated with multiple GPU and CPU intensive workloads with a diverse hardware utilization profile, including machine learning and microservice applications. The results of the evaluation were logically organized into two parts. First, the approximation of total system power by the Kepler framework demonstrated strong correlation with measured values, evidenced by an average regression coefficient of 1.01 and $R^2$ values of 0.95, indicating high predictive accuracy. Although deviations from the regression line were minimal, with a median of ~1.16 W, the ~5.23 W average static y-intercepts revealed inaccuracies in idle power consumption approximation. These findings suggest a promising potential of a simplified initial calibration procedure for y-intercepts to be statically applied within simulation scenarios without relying on external power meters. Nevertheless, certain machine learning workloads, both CPU and GPU based, showed higher deviation and lower $R^2$ values, indicating the need for workload-aware initial calibration depending on the desired accuracy. Secondly, in terms of per-process power calibration, Kepler's approximation for GPU workloads showed proportional scaling with system process power approximation, which is generally excluded from workload process approximation. Calibration improved by redistributing system process power consumption among active processes based on their power share, resulting in a ~50% increase in accuracy for GPU workloads relative to Kepler's approximation. Conversely, CPU workloads demonstrated minimal need for calibration, with a modest accuracy improvement of ~3.5%. Notably, valuable power insights for all microservices are provided, which can be presented useful for e.g. energy-aware scheduling strategies and debugging purposes.

Instead of focusing on any calibration improvements, future work could be inspired by Kepler's model repository [66], which may be expanded to incorporate calibration data derived from streamlined benchmarks. This could potentially eliminate the need for an initial calibration phase and dependency on external power meters by leveraging results from analogous systems. Within the context of power measurement, it is acknowledged that the socket-meter used in the prototype of thesis is costly and achieves high accuracy and sampling rates. An inexpensive and less precise option would consequently result in a less accurate calibration. A reduced sampling rate, on the other hand, does not necessarily compromise overall accuracy but may fail to capture transient power variations like spikes or drops. To address this limitation, a viable methodology could include interpolating between data points and simulating these transient events through monitoring and adjusting based on the approximated power consumption behavior.

Another direction could include the issue of renewable energy production often providing a constrained amount of power, which results in a defined power budget in many carbon-aware strategies [18], [19], [91]–[93]. Computing nodes that share the same power budget would operate within this collective limitation, competing for the necessary resources to meet their individual power needs. The testbed's ability for resource management and orchestration of nodes/applications could be utilized to redistribute power budgets among nodes based on user-defined policies. For example, in Wiesner et al.'s *FedZero* [18] approach for distributed machine learning, power would be first given to clients below their minimum participation, based on energy need. Then, remaining power is given to clients below their maximum participation, also based on energy need. The idea is similar to Costero et al.'s approach [60] but is not limited to DVFS as discussed in Chapter 3. Recreating FedZero's evaluation with the testbed of this thesis using real computing nodes opposed to a full simulation could yield additional insights as full simulations typically do not accommodate the unpredictable nature of energy availability and spare capacity of these nodes [15], [18], [20], [21], [93]–[95], which often deviate from forecasts [96], [97] that are usually utilized for these models.

# Appendix A

# Additional Figures

This appendix contains the additional figures 19 and 20 that were not included in the main body of the thesis as they were deemed non-essential for the understanding of the architecture of the DSB workloads Hotel Reservation and Media Service.

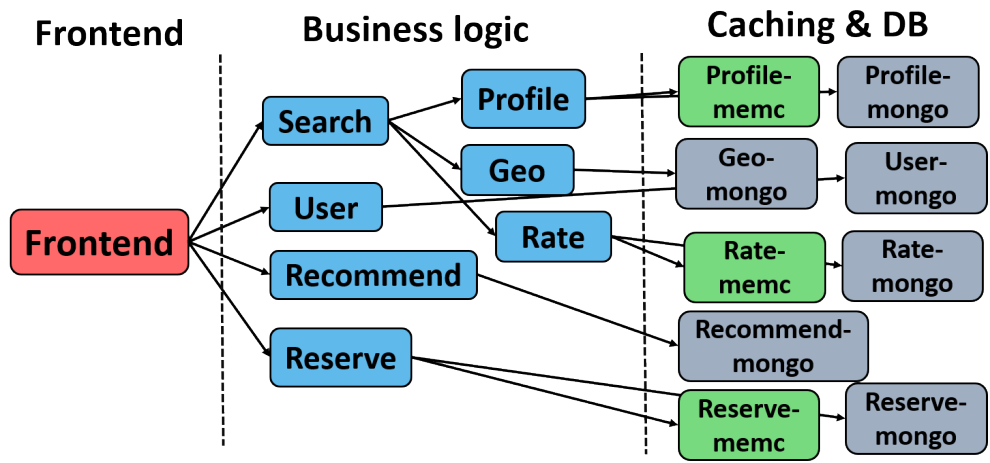

Figure 19: Microservices Dependency Graph for the Hotel Reservation Service [80].

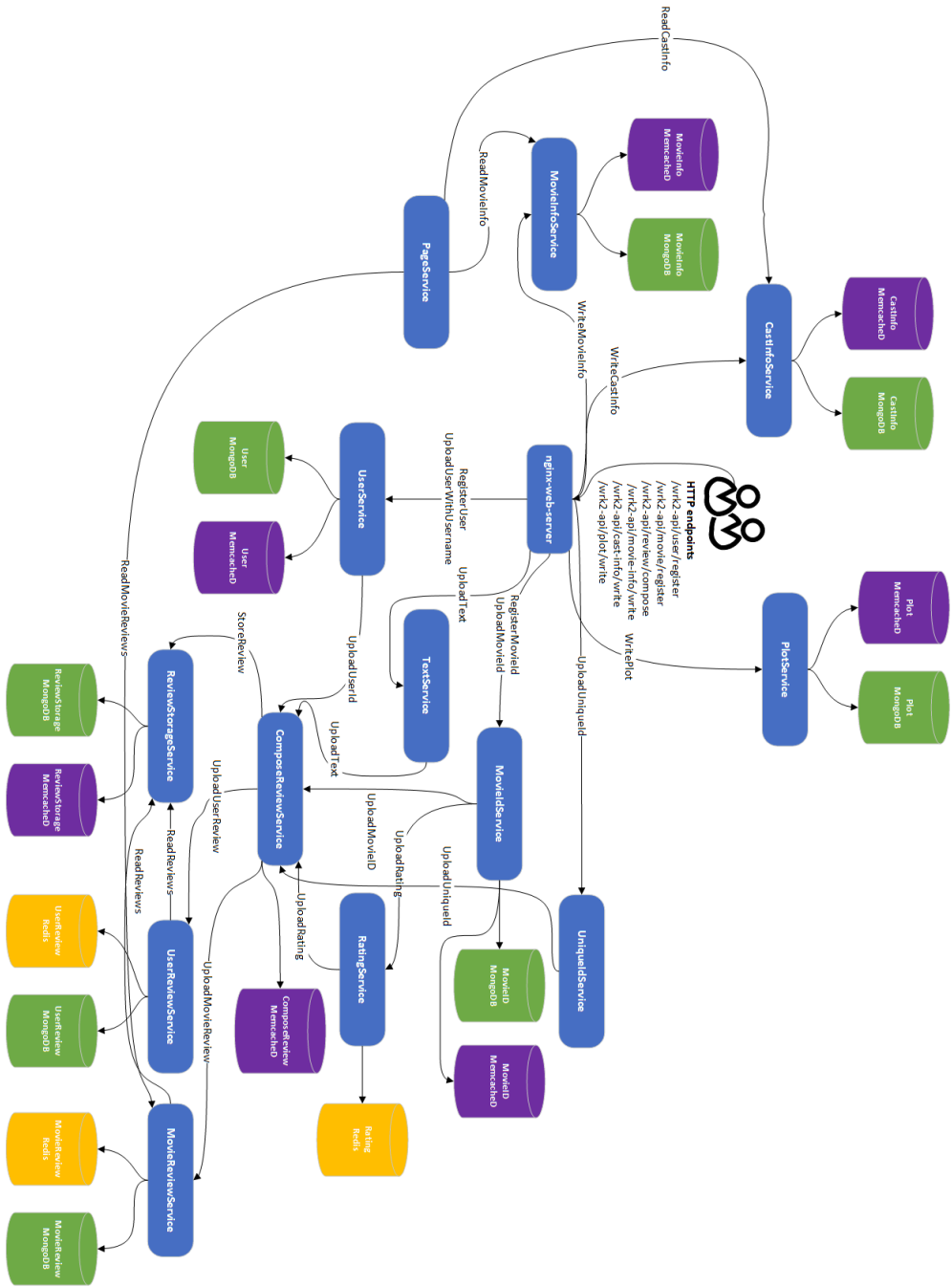

Figure 20: Microservices Dependency Graph for the Media Service [81]